\documentclass[pra,twocolumn,amsmath,amssymb,floatfix]{revtex4-1}
\usepackage{graphicx}% Include figure files
\usepackage{epsfig}% Include figure files
\usepackage{dcolumn}% Align table columns on decimal point
\usepackage{bm}% bold math
\usepackage{color}
\usepackage{titlesec}
\usepackage{cases}
\usepackage{cleveref}
\usepackage[normalem]{ulem}

\def\(({\left(}
\def\)){\right)}
\def\[[{\left[}
\def\]]{\right]}

\newcommand{\beq}{\begin{equation}}
\newcommand{\eeq}{\end{equation}}
\newcommand{\barr}{\begin{eqnarray}}
\newcommand{\earr}{\end{eqnarray}}
\newcommand{\bei}{\begin{itemize}}
\newcommand{\eei}{\end{itemize}}

\begin{document}
\title{\emph{Large-scale} thermalization, prethermalization and impact of the temperature\\ in the quench dynamics of two unequal Luttinger liquids}

\author{Paola~Ruggiero}
\affiliation{Department of Quantum Matter Physics, University of Geneva, 24 Quai Ernest-Ansermet, CH-1211 Geneva, Switzerland}
\author{Laura~Foini}
\affiliation{IPhT, CNRS, CEA, Universit\'{e} Paris Saclay, 91191 Gif-sur-Yvette, France}
\author{Thierry~Giamarchi}
\affiliation{Department of Quantum Matter Physics, University of Geneva, 24 Quai Ernest-Ansermet, CH-1211 Geneva, Switzerland}

\date{\today}

\begin{abstract}
We study the effect of a quantum quench between two tunnel coupled Tomonaga-Luttinger liquids (TLLs) with different speed of sound and interaction parameter.
The quench dynamics is induced by switching off the tunnelling and letting the two systems evolve independently. We fully diagonalize the problem within a quadratic approximation for the initial tunnelling. 
Both the case of zero and finite temperature in the initial state are considered. We focus on correlation functions associated with the antisymmetric and symmetric combinations of the two TLLs (relevant for interference measurements), 
which turn out to be coupled due to the asymmetry in the two systems' Hamiltonians.
The presence of different speeds of sound leads to multiple lightcones separating different decaying regimes. In particular, in the large time limit, we are able to identify a prethermal regime where the two-point correlation functions of vertex operators of symmetric and antisymmetric sector can be characterized by two emerging effective temperatures, eventually drifting towards a final stationary regime that we dubbed \emph{quasi-thermal}, well approximated at large scale by a thermal-like state, where these correlators become time independent and are characterized by a unique  correlation length. If the initial state is at equilibrium at non-zero temperature $T_0$, all the effective temperatures acquire a linear correction in $T_0$, leading to faster decay of the correlation functions. Such effects can play a crucial role for the correct description of currently running cold atoms experiments.
\end{abstract}

\maketitle

%\textcolor{red}{\tableofcontents}

\section{Introduction}

The out-of-equilibrium physics of low dimensional many-body quantum systems has witnessed important theoretical advances in recent times \cite{Polkovnikov_ColloquiumNonEquilibrium, Cazalilla_DynamicsThermalization, Cazalilla_RevUltracold, Calabrese_RevQuenches, Calabrese_IntroIntegrabilityDynamics, Gogolin_ReviewIsolatedSystems, DAlessio_ETH,Abanin_ColloquiumMBL}. Several long-standing questions about the relaxation dynamics and phenomena like equilibration, thermalization, emergence of statistical mechanics from microscopics \cite{Deutsch_ETH, Srednicki_ETH, Deutsch_OriginThermodynamicEntropy, Alba_EntanglementThermodynamics, Calabrese_QuenchesCorrelationsPRL}, as well as lack or generalized forms of thermalization
have been addressed both in clean and disordered models~\cite{Basko_ConjectureMBL, Gornyi_ConjectureMBL, Giamarchi_MBL,Rigol_RelaxationHardCoreBosons, Rigol_GGE, Bertini_GHD, CastroAlvaredo_GHD}.
Remarkably, a large number of such predictions have been confirmed in cold atoms experiments~\cite{Bloch_RevUltracold,Langen_RevUltracoldAtoms}, which allowed to engineer quantum many-body Hamiltonians reproducing models of theoretical interest~\cite{Gorlitz_RealizationBoseCondensate,Greiner_RealizationBoseCondensate,Kinoshita_ObservationTonksGirardeau,Kinoshita_NewtonCradle,Hofferberth_DynamicsBoseGases,Trotzky_Relaxation1dBoseGas,Cheneau_ExperimentalLightcones,Gring_Prethermalization,Langen_ThermalCorrelationsIsolatedSystems,Langen_ExperimentGGE,Langen_PrethermalizationNearIntegrable,Kaufman_ThermalizationViaEntanglement, Schweigler_NonGaussianCorrelations, Schemmer_ExperimentGHD}.

\begin{figure}[t]
  \begin{center}
        \includegraphics[height=5.cm]{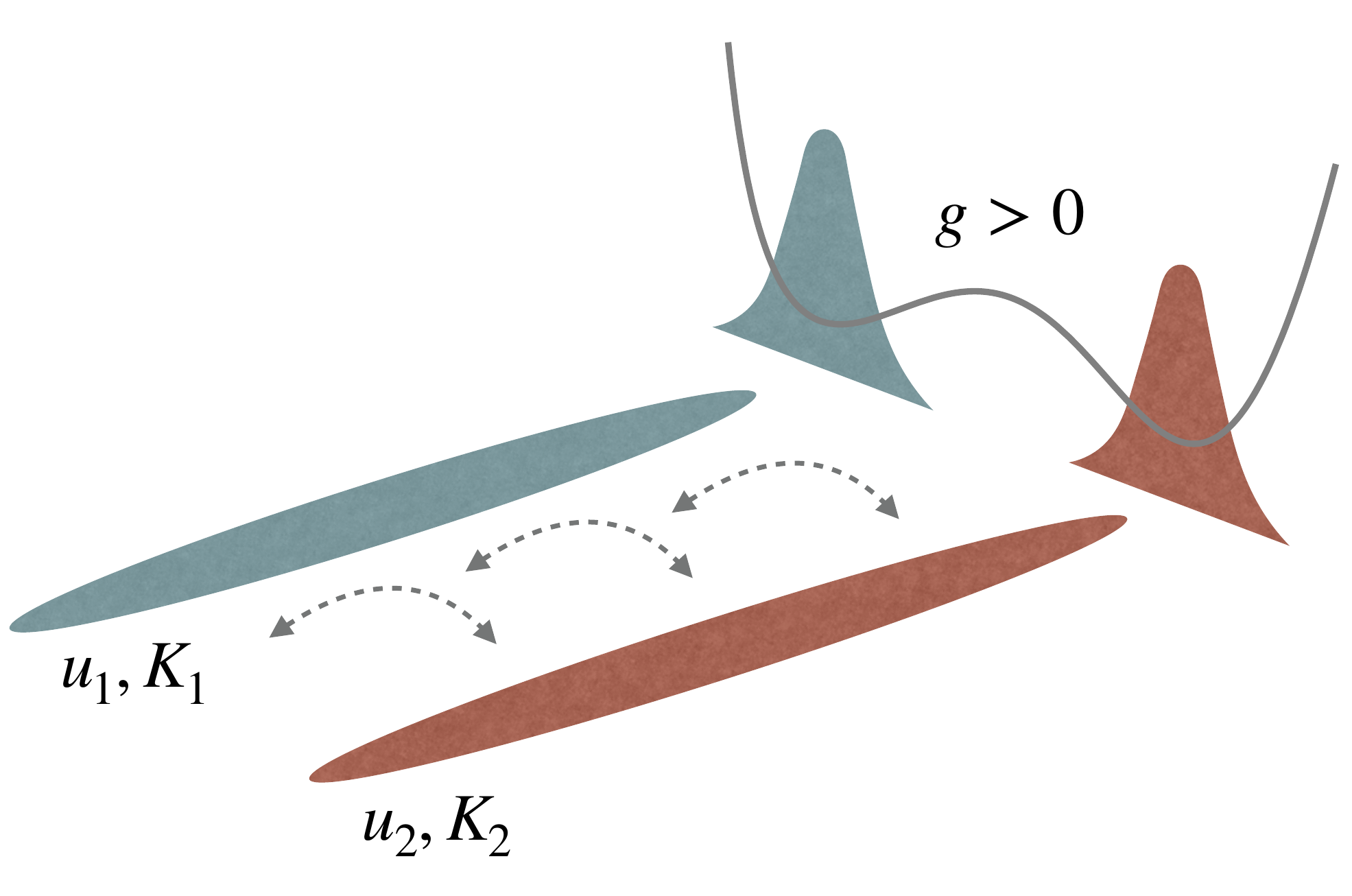}
                 \caption{Schematic picture of the system studied in this paper. It consists of two unequal Luttinger liquids, with sound velocities $u_i$ and Luttinger parameters $K_i$ ($i=1,2$). We quench the system by switching off the tunnelling $g$, starting from a non zero value (both ground state and finite temperature thermal states are considered as initial states). This corresponds to suddenly raising the barrier of the double well potential separating the two sides.
        }\label{cartoonLL}
  \end{center}
\end{figure}

Among the different experimental setups, an interesting example is offered by matter-wave interferometry \cite{Andrews_FirstExperimentInterferenceCondensates}, using pairs of split one-dimensional Bose gases \cite{Shin_ExperimentSplitCondensates1, Shin_ExperimentSplitCondensates2, Shin_ExperimentSplitCondensates3, Schumm_ExperimentMatterWaveInterferometry, Albiez_ObservationTunnellingBosonicJosephsonJunction,Gati_RealizationBosonicJosephsonJunction, Levy_ExperimentalBosonicJosephsonEffects, Kuhnert_ExpEmergenceCharacteristicLength1d}.
Effectively, such systems consist of two tunnel-coupled one-dimensional (1d) interacting tubes, whose low-energy physics maps to a pair of indepependent TLLs~\cite{Tomonaga_TLLiquid,Luttinger_TLLiquid,Haldane_LuttingerLiquid,giamarchi2004}, plus a coupling resulting from the tunnelling (a schematic representation is given in Fig.~\ref{cartoonLL}).

In the theoretical description, it is often assumed that the two TLLs are \emph{identical}, meaning they are characterized by equal sound velocities and Luttinger parameters.
In this case, the theory consists of a quantum sine-Gordon model and a free boson~\cite{Kardar_FermionicJosephsonJunctionAsSineGordon,Gritsev_LinearResponseCoupledCondensates}, describing respectively the antisymmetric and symmetric combinations of the phase fields (see section \ref{Sec_model} for proper definitions). Importantly, as a consequence of the symmetry between the two TLLs, these two sectors are not coupled and thus can be treated as isolated systems.
In particular, time-dependent correlation functions of the antisymmetric sector (directly related to interference measurements~\cite{Polkovnikov_InterferenceCondensates}) after a sudden change in the tunnelling strength (a so-called quantum quench~\cite{Calabrese_QuenchesCorrelationsPRL}) have been widely studied~\cite{Imambekov_LectureNotesMatterInterferometry}. They have been obtained by relying, for example, on a simple harmonic approximation~\cite{Iucci_QuenchLL,Iucci_QuenchSineGordon,Foini_CoupledLLsMassiveMassless,Foini_CoupledLLSchmiedtmayerMassive} and, more recently, on a refined selfconsistent version of it~\cite{vanNieuwkerk_QuenchSineGordonSelfConsistentHarmonicApprox,vanNieuwkerk_TunnelCoupledBoseGasesLowEnergy}. Exact results have been further obtained at the Luther--Emery point~\cite{Iucci_QuenchSineGordon}, by means of techniques such as integrability~\cite{Bertini_QuenchSineGordon,Cubero_QuenchAttractiveSineGordon,Gritsev_LinearResponseCoupledCondensates} and semi-classical methods~\cite{Kormos_SemiclassicalQuantumSG,Moca_SemiclassicalQuantumSG}. A truncated conformal approach was considered in~\cite{Kukuljan_QuenchSineGordonTruncatedConformalApproach,Horvath_TruncatedApprox}, while a combination of analytic (based on Keldysh formalism~\cite{kamenev2011field}) and numerical methods was used in \cite{DallaTorre_QuenchSineGordonMasslessMassive}.
Finally, an effective model for the relative degrees of freedom was recently derived in~\cite{Tononi_DynamicsTunnellingQuasicondensates}. In these studies the existence of a \emph{prethermal} regime was demonstrated. 

Much less attention has been devoted so far to the effect of introducing an ``imbalance'' between the two systems. On the theory side such a case is interesting since, due to the presence of two velocities, one can expect multiple lightcones to emerge, separating different decaying regimes (as opposed to the single lightcone effect~\cite{Calabrese_QuenchesCorrelationsPRL,Cheneau_ExperimentalLightcones} usually observed in systems of identical TLLs~\cite{Foini_CoupledLLSchmiedtmayerMassive,Langen_ThermalCorrelationsIsolatedSystems}). 
Because of the coupling between the modes one can also expect that the prethermal regime evidenced in the antisymmetric sector to decay into another final regime. 
{Whether such a regime could be characterized by a single temperature despite the integrable nature of the underlying model~\cite{Rigol_RelaxationHardCoreBosons, Rigol_GGE} is an interesting question.
However due to the complexity of such a situation the asymmetric case has been much less studied. Noteworthy exceptions are provided by Ref.~\cite{Langen_UnequalLL,Kitagawa_DynamicsPrethermalizationQuantumNoise}, where, relying on a phenomenological approach for the quench (especially concerning the initial state), the authors consider different forms of imbalance for two examples of systems described by LLs.

Given the importance of the physical effects in the asymmetric situation, it would thus be highly desirable: 
i) to have a full theoretical derivation of the quench of two different TLLs; 
ii) to allow for all possible sources of imbalance between them and disentangle the effects coming from unequal sound velocities $u_i$ from the ones related to different Luttinger parameters $K_i$ ($i=1,2$). Such a study is the goal of the present paper.

The paper is organized as follows. In section~\ref{Sec_model} we introduce the model and the quench dynamics we focus on.
Section~\ref{sec:Bog} and section \ref{sec:correlations} discuss the Bogoliubov transformation which diagonalize the hamiltonian at initial time and introduce the correlation functions of interest, respectively.
In section~\ref{sec:GS} a detailed analysis of the dynamics when starting from the ground state (i.e., at zero temperature) of the initial hamiltonian is carried out. The same analysis is extended to quenches starting from a thermal states in section~\ref{Sec_thermal_state}.
A discussion of the results, also in connection with previous literature, is left to section \ref{sec:discussion}.
Conclusions and future perspectives are finally collected in section \ref{sec:conclusions}.
Details regarding the calculations are reported in the appendices.

\section{Setting of the quench}\label{Sec_model}

We consider two different Luttinger liquids which are initially tunnel-coupled and then evolve
independently: this is one of the simplest situation one can look at, since the evolution is the one of two free (compactified) bosons, while the coupling between the two is only in the initial state.
This protocol has also the advantage to be easily implementable in a controlled way in cold atom experiments. 

Microscopically, the system corresponds to two interacting 1d Bose gases, represented by bosonic fields $\Psi_i$ ($i=1,2$) of mass $M_i$ and short-ranged two-body interactions that can be represented by a delta function of strength $\mathcal{U}_{i}$.
We are going to work with their phase $\theta_i(x)$ and the fluctuation of the densities $n_i(x)$, related to the original field via the bosonization formula~\cite{Haldane_LuttingerLiquid,Cazalilla_RevUltracold,giamarchi2004}
\begin{equation}
\Psi_i (x)= \sqrt{\rho_i+n_i(x)} e^{i\theta_i(x)} \ ,
%, \quad n_i (x) =\Psi_i^{\dagger} (x) \Psi_i (x)-\rho_i.
\end{equation}
with $[n_i(x),\theta_j(x')] = i \hbar \delta(x-x') \delta_{i,j}$ and $\rho_i$ is the average density of the $i$-th tube.
In terms of these variables, the system is supposed to be prepared in the ground state (or in a thermal state) of the
(generalized) Sine-Gordon Hamiltonian
\beq\label{H_initial}
H_{\text{initial}}^{\text{SG}} = H_1 + H_2 - \frac{g}{2\pi} \int {\rm d} x \ \cos(\theta_1(x)-\theta_2(x)),
\eeq\
where $H_i$ are the Luttinger liquid Hamiltonians~\cite{giamarchi2004}
\beq\label{H_LL}
H_i =  \frac{\hbar}{2 \pi} \int {\rm d} x \ \left[ u_i K_i (\nabla \theta_i)^2 + \frac{u_i}{K_i} (\pi n_i)^2 \right] \ ,
\eeq
and the cosine term originates from the tunnelling ($\Psi_1^{\dagger} \Psi_2 +\text{h.c.}$), with strength tuned by $g$.
In (\ref{H_LL}) $K_i$ is the Luttinger liquid parameter which encodes
the interaction of the system and $u_i$ is the speed of sound. They are related to the microscopic parameters. 
Such relations are known analytically in the weak interaction regime
\beq\label{micro_parameters}
K_i= \hbar \pi \sqrt{\frac{\rho_i}{M_i \mathcal{U}_i}} \ , \quad u_i=\sqrt{\frac{\mathcal{U}_i \rho_i}{M_i}} \ ,
\eeq
and can be extracted numerically otherwise \cite{Cazalilla_RevUltracold}.
Therefore one can get unequal TLLs in many different settings, depending on the values of $M_i, \mathcal{U}_i$ and $\rho_i$.

Hereafter we will set $\hbar=1$.
At time $t=0$ the interaction between the two systems is switched off
and the final Hamiltonian simply reads
\beq\label{H_final}
H_{\text{final}} = H_1 + H_2 \ .
\eeq
As the study of the initial hamiltonian (\ref{H_initial}) is particularly involved, we resort to a semiclassical (harmonic) approximation
\beq\label{Semiclassical}
H_{\text{initial}}^{\text{SC}} = H_1 + H_2 + \frac{g}{4\pi} \int {\rm d} x \ (\theta_1(x)-\theta_2(x))^2 \ .
\eeq
Note that in our quench the approximation is only in the initial state, while the dynamics can be obtained exactly.
Such approximation is expected to hold as long as the cosine term in \eqref{H_initial} is highly
relevant in a renormalization group (RG) sense (in the case of identical TLLs, this corresponds to $K$ large enough~\cite{giamarchi2004}, while the same RG analysis is missing for the more generic case considered here;
note, however, that in the experiments involving bosons with contact interactions we can safely assume that we are in the relevant regime).
Remarkably, for identical TLLs, it has been shown by means of exact calculations that
the dynamics starting from (\ref{Semiclassical}) is qualitatively the same as from the Luther--Emery point
where the full cosine term can be taken into account~\cite{Iucci_QuenchSineGordon}.

The fields $\theta_i$ and $n_i$ admit a decomposition in normal
modes
\beq
\begin{array}{ll}\label{expansion_theta}
\displaystyle  \theta_i(x) = &  \displaystyle \frac{i}{\sqrt{L}}  \sum_{p \neq 0} \  e^{ - i p x}
\sqrt{\frac{\pi}{2 K_i |p|}} (   b^{\dag}_{i,p} -b_{i,-p})
\displaystyle + \frac{1}{\sqrt{L}} \theta_{i,0}
\end{array}
\eeq
\beq
\begin{array}{ll}
 \displaystyle n_i(x) = &  \displaystyle \, \frac{1}{\sqrt{L}}  \sum_{p\neq 0} e^{ - i p x}
\sqrt{\frac{|p| K_i}{2 \pi}} (b^{\dag}_{i,p} +  b_{i,-p} )
\displaystyle + \frac{1}{\sqrt{L}} n_{i,0} \ .
\end{array}
\eeq
 where $L$ is the system size. In the rest of the paper we will only focus on the thermodynamic limit (TDL), namely infinite system size. Finite-size effects will be discussed elsewhere.
In terms of these bosons the final Hamiltonian is diagonal, namely
\beq
H_{\text{final}} = u_1 \sum_{p\neq 0} |p| b^{\dag}_{1,p} b_{1,p} +  u_2 \sum_{p\neq0} |p| b^{\dag}_{2,p} b_{2,p} \ ,
\eeq
where the zero modes (i.e., $\theta_{i,0},n_{i,0} $) have been neglected.
The Hamiltonian $H_{\text{initial}}^{\textrm{SC}}$, instead, is quadratic but needs to be diagonalized via a Bogoliubov transformation (see Section \ref{sec:Bog} below).

To highlight the difference with the case of two identical systems
it is useful to introduce the symmetric ($+$) and antisymmetric ($-$) modes
\beq
\begin{array}{ll}
\displaystyle
%\theta_{S\char`\\  A}(x)
\theta_{\pm} (x)= \frac{1}{\sqrt{2}} (\theta_1(x)\pm \theta_2(x))
\\ \vspace{-0.2cm} \\
 \displaystyle
 %n_{S\char`\\  A}(x)
 n_{\pm} (x) =\frac{1}{\sqrt{2}}  (n_1(x)\pm n_2(x) )
 \end{array}
\eeq
which satisfy canonical commutation relations.
In terms of these variables the final Hamiltonian reads
\begin{multline}\label{H_piu_meno}
H_{\text{final}} =
\displaystyle \frac{1}{2\pi} \int {\rm d} x \left\{ u K \left[  (\nabla \theta_+)^2 + (\nabla \theta_-)^2 \right] \right. \\
\left. \qquad \qquad + \frac{u}{K} \left[  (\pi n_+)^2 + (\pi n_-)^2 \right] \right\}\\
+   \frac{1}{\pi} \int {\rm d} x \left\{
\Lambda_{\theta}  \nabla \theta_+ \nabla\theta_- + \Lambda_{n} \pi^2 n_+ n_-
\right\}
\end{multline}
with
\beq \label{Lpar_pm}
\frac{u}{K}  = \frac12 \left( \frac{u_1}{K_1} + \frac{u_2}{K_2} \right)
\quad
u K =  \frac12 \left( u_1 K_1 + u_2 K_2 \right)
\eeq
\beq \label{lambdas}
\Lambda_{\theta} =   \frac12 \left( u_1 K_1 - u_2 K_2 \right)
\quad
\Lambda_{n} =  \frac12   \left( \frac{u_1}{K_1} - \frac{u_2}{K_2} \right) \ .
\eeq
Therefore we see that in the case of two identical systems the final
hamiltonian dispay decoupling between symmetric and
antisymmetric sectors and the quench occurs only in the antisymmetric one.

The situation that we consider in this work is more involved as this decoupling is
not possible and to study correlation functions of $\theta_{-}$, which are usually those
of experimental interest, one has to consider the dynamics of $\theta_1$
and $\theta_2$ which are correlated via the initial condition.

\section{Bogolioubov transformation for two species of bosons} \label{sec:Bog}

In order to characterize the evolving state we aim at diagonalizing
the initial Hamiltonian $H_{\text{initial}}^{\textrm{SC}}$ and write it as
\beq\label{Initial_diagonal}
H_{\text{initial}}^{\text{SC}} = \sum_{p\neq0} \lambda_{m,p} \eta_{m,p}^{\dag} \eta_{m,p} + \sum_{p\neq0} \lambda_{0,p} \eta_{0,p}^{\dag} \eta_{0,p} \ ,
\eeq
up to an unimportant overall constant, which we neglect.
The meaning of the subscripts $m,0$ will be clearer in the following: they emphasize that, as we are going to show, the two diagonal modes above are \emph{massive} ($m$) and \emph{massless} ($0$) respectively.

The transformation bringing the hamiltonian in the form \eqref{Initial_diagonal} amounts to a Bogoliubov
rotation of a four component vector, mixing the modes $(p,-p)$ of the two initial species of bosons.
Specifically, we introduce the vectors of bosons of the initial and the final Hamiltonian,
 $\boldsymbol{\eta}^{\dag}_p=(\eta_{m,p}^{\dag} \, \eta_{m,-p} \, \eta_{0,p}^{\dag} \, \eta_{0,-p} )$
and $\boldsymbol{b}^{\dag}_p =(b_{1,p}^{\dag} \, b_{1,-p} \, b_{2,p}^{\dag}  \, b_{2,-p})$.
These two are related by a matrix multiplication $\boldsymbol{b}_p = B(\hat\varphi_p)\boldsymbol{\eta}_p$
with $B(\hat\varphi_p)$ depending on the set of parameters
 $\hat\varphi_p=\{\varphi_{1,p},\varphi_{2,p},\Delta_p,\phi_p\}$ and parametrized as follows \cite{Elmfors_Bogoliubov}
\beq\label{Bogoliubov_matrix}
B(\hat\varphi_p)  =
\begin{bmatrix}
B_2(\varphi_{1,p}) \cos\phi_p & B_2(\varphi_{2,p}-\Delta_p) \sin\phi_p
\\ \vspace{-0.2cm} \\
- B_2(\varphi_{1,p} +\Delta_p)\sin\phi_p & B_2(\varphi_{2,p}) \cos\phi_p
\end{bmatrix}
\eeq
with
\beq
B_2(\varphi) = \left[
\begin{matrix}
\cosh\varphi & \sinh\varphi
\\ \vspace{-0.2cm} \\
\sinh\varphi & \cosh\varphi
\end{matrix}
\right] .
\eeq\\
Details on the derivation are reported in Appendix~\ref{appendix:Bogoliubov}.
The parameters of the matrix $B(\hat\varphi_p) $ in (\ref{Bogoliubov_matrix}) have the following interpretation: $\varphi_{1,p}$ and $\varphi_{2,p}$ define Bogoliubov rotations associated to the two bosons, separately. $\phi_p$ is the mixing angle between them. Finally, $\Delta_p$ exists only when the Bogoliubov rotation and the mixing of different bosons appear at the same time \cite{Elmfors_Bogoliubov}.
Explicitly, they are given by
\begin{widetext}
\begin{eqnarray}
\begin{array}{c}
\displaystyle \nonumber
\varphi_{1,p} = \frac12 \log \left( \frac{\lambda_{m,p}}{u_1 |p|} \right)
% \\ \vspace{-0.2cm} \\
\displaystyle
\qquad \varphi_{2,p} = \frac12 \log \left( \frac{\lambda_{0,p}}{u_2 |p|} \right)
%\\ \vspace{-0.2cm} \\
\displaystyle
\qquad \Delta_p = \Delta = \frac12 \log \left( \frac{u_1}{u_2} \right)
\end{array}\\
%\eeq
%
%\beq
\label{parameters}
\phi_p = \text{arctan}\left[
\frac{  \sqrt{  \epsilon_{1,p}^2 -  (u_1p)^2 } \sqrt{ \epsilon_{2,p}^2 - (u_2 p)^2  }}{
       ( \epsilon_{1,p}^2 -  \epsilon_{2,p}^2 )+
     \sqrt{
     ( \epsilon_{1,p}^2 +  \epsilon_{2,p}^2)^2 -
        \left[ (u_2 p)^2  \epsilon_{1,p}^2 +   (u_1 p)^2  \epsilon_{2,p}^2 -
         (u_1p)^2 (u_2 p)^2\right]}  } \right] \,
\end{eqnarray}
in terms of (for $i=1,2$)
\beq
\begin{array}{l}
\displaystyle
\epsilon_{i,p} = \sqrt{u_i |p| \left( u_i |p|+ \frac{g}{2 K_i |p|} \right)} ,
\end{array}
\eeq
and the eigenvalues of the hamiltonian (\ref{Initial_diagonal}) (for $k=m,0$) 
\beq \label{Eigenvalues}
\lambda_{k,p} =
\frac{1}{\sqrt{2}} \sqrt{
   \epsilon_{1,p}^2 +  \epsilon_{2,p}^2  \pm
   \sqrt{ ( \epsilon_{1,p}^2 +  \epsilon_{2,p}^2)^2 -
     4 \left[  (u_2 p)^2  \epsilon_{1,p}^2 + (u_1 p)^2 \epsilon_{2,p}^2 - (u_1 p)^2  (u_2 p)^2\right]
     }  } \ .
\eeq
\end{widetext}
where the $+ \, (-)$ sign is associated to the $m \, (0)$-mode. 
Note that, at the leading order in $p\to 0$, the eigenvalues \eqref{Eigenvalues} read
\beq\label{Eig_p0}
\lambda_{m,p} = m_0 \ ,  \qquad \qquad \lambda_{0,p} = a |p| \ , 
\eeq
with
\beq
m_0 = \sqrt{\frac{g u}{K}} \ ,  \qquad\qquad a = \sqrt{\frac{u_1 u_2}{K_1K_2}} K \,
\eeq
in terms of the parameters in (\ref{Lpar_pm}). Therefore, as anticipated, they describe a massive
and a massless mode.
Note also that, in the limit of equal TLLs, they would coincide with the antisymmetric and the symmetric modes,  respectively.

\section{Correlation functions after the quench} \label{sec:correlations}

We will be mostly interested in the correlation functions of vertex operators
\begin{multline} \label{eq:Cpm}
C_{\pm}(x,t,T_0) \equiv
%& \displaystyle
 \langle  e^{i \sqrt{2} [ \theta_{\pm}(x,t) - \theta_{\pm}(0,t)]}  \rangle_{T_0}= \\
= e^{- \langle [ \theta_{\pm}(x,t) - \theta_{\pm}(0,t)]^2 \rangle_{T_0}} \ ,
\end{multline}
where the expectation value $\langle \cdot \rangle_{T_0}$ is on the initial state, which we choose to be either the ground state ($T_0=0$) or a finite temperature ($T_0\neq 0$) equilibrium state of the initial hamiltonian ${H_{\text{initial}}^\text{SC}}$.

While the function $C_- (x, t, T_0)$ is of clear experimental relevance and has been directly measured using matter-wave interferometry \cite{Kuhnert_ExpEmergenceCharacteristicLength1d,Langen_ThermalCorrelationsIsolatedSystems, Langen_ExperimentGGE,Gring_Prethermalization}, observables within the symmetric sector as $C_+ (x, t, T_0)$ have not been measured so far.
Nonetheless, very recently, it was pointed out that that correlation functions in
the symmetric sector also contribute to the measured density after
time-of-flight \cite{vanNieuwkerk_ProjectivePhaseMeasurements}, thus giving hopes for their future measurements.

Note that in our approach, due to the absence of decoupling between
symmetric and antisymmetric variables, $\theta_{\pm}$ are not anymore the preferable variables to work with (as it was the case in the symmetric quench~\cite{Foini_CoupledLLsMassiveMassless,Foini_CoupledLLSchmiedtmayerMassive}).
Instead, we will stick to the initial fields, $\theta_1$ and $\theta_2$. In terms of those variables, the
one (two) point function of the symmetric or antisymmetric fields is recast
into a two (four) point function.

We start by defining the parameters $\hat\epsilon_p=\{u_1|p|,u_2|p|\}$ entering in the definition of the time evolution operator
\beq
\displaystyle \label{eq:Ut}
U(\hat\epsilon_p,t) \equiv
\begin{bmatrix}
e^{- i u_1 |p| t}  & 0 & 0 & 0 \\
0 & e^{ i u_1 |p| t}  & 0 & 0 \\
0 & 0 & e^{- i u_2 |p| t}  & 0 \\
0 & 0 & 0 & e^{ i u_2 |p| t}
\end{bmatrix}
\eeq
and the matrices
\beq
\begin{array}{ll}
\displaystyle
P^{\pm} \equiv  &
\displaystyle
\sum_{ij}
\frac{(\pm1)^{i+j}}{4 K_i K_j}
\begin{bmatrix}
\delta_{i1}\delta_{j1} & \delta_{i1}\delta_{j2}
\\ \vspace{-0.2cm} \\
 \delta_{i2}\delta_{j1} & \delta_{i2}\delta_{j2}
 \end{bmatrix}
\otimes
\begin{bmatrix}
 1 & -1
\\ \vspace{-0.2cm} \\
 -1 & 1
 \end{bmatrix}
\end{array}
\eeq
where $\otimes$ denotes the Kronecker product.

For a generic quench starting from a thermal state of (\ref{Semiclassical})
at temperature $T_0$, Eq.~\eqref{eq:Cpm} takes the compact form (see Appendix~\ref{sec:Wpm} for details)
\beq\label{General_corr}
\begin{array}{ll}
\displaystyle
C_{\pm}(x,t,T_0)  \displaystyle = \exp\left[-  \int_0^\infty {\rm d} p \ e^{-\alpha^2 p^2}  \frac{2}{p} (1-\cos p x) \right.
\\ \vspace{-0.2cm} \\
 \displaystyle \qquad
 \Big(
W_{22}^{\pm} \text{cotanh}\left( \frac{\lambda_{m,p}}{2 T_0}\right)
%\\ \vspace{-0.2cm} \\
 \displaystyle
\left. + W_{44}^{\pm} \text{cotanh}\left( \frac{\lambda_{0,p}}{2 T_0}\right)  \Big)
\right]\ .
\end{array}
\eeq
For convenience we have introduced an ultraviolet cutoff $\alpha^{-1}$.
We further denoted by $W^{\pm}_{\mu \nu}$ the elements of the matrices
\beq \label{eq:Wpm}
W^{\pm} \equiv B^\dag(\hat\varphi_p)  U^\dag (\hat\epsilon_p,t) P^{\pm} U(\hat\epsilon_p,t) B(\hat\varphi_p) \ .
\eeq
Note that only two elements of the whole matrices are needed to fully characterize the correlation functions \eqref{General_corr}. Moreover, thanks to the quadratic approximation in the initial hamiltonian, they can be written explicitly (see Eq.~\eqref{W2244} in Appendix~\ref{sec:Wpm}).

In order to define effective temperatures for $C_{\pm}$, we are going to compare these post-quench correlations with the equilibrium ones
at finite temperature $T$
\beq\label{Eq_Cpm_eq}
\begin{array}{l}
\displaystyle
C_{\pm}^{\text{eq}}(x,t,T) = \exp\left[ -  \int_0^\infty {\rm d} p \ e^{-\alpha^2 p^2}  (1-\cos p x) \frac{1}{2 p}  \right.
\\ \vspace{-0.2cm} \\
 \displaystyle \qquad
\left.
\times \left[ \frac{1}{K_1} \text{cotanh}\left( \frac{u_1 |p|}{2 T}\right)  + \frac{1}{K_2} \text{cotanh}\left( \frac{u_2|p|}{2 T}\right)
\right]
\right]
\end{array}
\eeq
which present an exponential decay in space with (inverse) correlation length
\beq\label{Eq_corr_length}
\xi_{T}^{-1} = \frac{\pi}{2}  \left( \frac{1}{u_1 K_1}+\frac{1}{u_2 K_2} \right) T \ .
\eeq
%\textcolor{blue}{The meaning of such effective temperatures will be carefully discussed within the paper.}

\section{Quench from the ground state} \label{sec:GS}

We consider here the quench from the ground state ($T_0 = 0$) of the Hamiltonian (\ref{Semiclassical})
and we defer the solution of the dynamics from a thermal state at temperature $T_0$
to section \ref{Sec_thermal_state}. In this section, expectation values over the ground state will be simply denoted as $\langle \cdot \rangle$.

\subsection{Eigenmodes dynamics} \label{sec:eigdynamics}
An important observation is that in the limit $T_0 \to 0$, we have $\text{cotanh}(\lambda_{k,p} /(2T_0)) \to1$ in Eq. (\ref{General_corr}),
and it turns out that the leading order as $p\to0$ of $C_{\pm}$ is captured uniquely by the first term, namely by the massive mode.
The main contribution is better characterized by introducing the
dynamics of the modes of phase and density. In particular, by using the following decomposition
\beq
\begin{array}{ll}
\displaystyle
\theta_i (x,t) = \sum_p e^{-i p x} \theta_{i}(p,t)  \ ,
\end{array}
\eeq
one finds that
\beq \label{theta_i_t}
\begin{array}{ll}
\displaystyle
\theta_i (p,t)
=   \cos(u_{i}|p| t) \theta_{i}(p,0) - \alpha_{i, p} \sin(u_{i}|p| t) n_{i}(p,0) 
\end{array}
\eeq
with $\alpha_{i, p}=\frac{\pi}{K_i |p|}$.
The expectation value of the two point function over the ground state simplifies to
\begin{multline}\label{Eigenmodes_dynamics}
\langle \theta_{i}(p,t) \theta_{j}(-p,t) \rangle = \\
 \cos(u_{i}|p| t) \cos(u_{j}|p| t) \langle \theta_{i}(p,0) \theta_{j}(-p,0) \rangle +\\
+ \sin(u_{j}|p| t) \sin(u_{j}|p| t)  \alpha_{i p}  \alpha_{j -p}  \langle n_{i}(p,0) n_{j}(-p,0) \rangle
\end{multline}
for $i,j=1,2$, namely the initial correlations between $\theta_i$ and $n_j$ do not enter in the eigenmodes' dynamics.

By plugging in the asymptotic expressions (\ref{Eig_p0}), one can further check
that at the leading order in $p\to0$ it holds
\beq\label{Expectation_modes_t0_1}
 \langle \theta_{i}(p,0) \theta_{j}(-p,0) \rangle \simeq \frac{\pi}{4 a |p|} \frac{u_1 u_2}{K_1 K_2} \frac{K}{u}
\eeq
\beq\label{Expectation_modes_t0_2}
 \langle  n_{i}(p,0) n_{j}(-p,0)   \rangle \simeq  \frac{(-1)^{i+j}}{2 \pi}  \sqrt{\frac{K_i K_j m_i m_j}{u_i u_j}} \ ,
\eeq
where we defined $m_1=m_0 \cos\phi_0^2 $, $m_2=m_0 \sin\phi_0^2$ and $\phi_p \simeq \phi_0=\arctan\sqrt{\frac{K_1}{K_2} \frac{u_2}{u_1}} $.
Note the initial anticorrelations between the densities
of the two systems, which will have a role on the evolution of the phase.

To evaluate Eq.~\eqref{General_corr} at large scale and times, the strategy is to proceed order-by-order in powers of $p$, which successively lead to exponential and power-law decay of correlations.
The leading divergence as $p\to 0$ in the integrand of $C_{\pm}(x,t) \equiv C_{\pm} (x,t,T_0=0)$ (cfr. Eq.~\eqref{General_corr})
comes from the initial density fluctuation while the part
coming from the phase is negligible (this is due to the term $\alpha_{i,p} \alpha_{j,-p} \propto 1/p^2$ in the eigenmodes' dynamics (\ref{Eigenmodes_dynamics})). Notice that since the sound velocity $a$ appears only in the phase fluctuations, at this order the massless mode will not play any role in the correlation functions, consistently with what anticipated from Eq.~\eqref{General_corr}.

If we define the building block of the correlations \eqref{General_corr} as
\begin{equation} \label{building_blocks}
c_{ij} (x,t) \equiv \langle(\theta_{i}(x,t) - \theta_{i}(0,t) ) (\theta_{j}(x,t) - \theta_{j}(0,t)) \rangle
\end{equation}
such that
\beq
\ln C_{\pm} =\frac{1}{2} \left(c_{11} + c_{22} \pm 2 c_{12} \right),
\eeq
then from \eqref{Eigenmodes_dynamics} and \eqref{Expectation_modes_t0_2} we have
\begin{multline} \label{ansatz}
%\langle [ \theta_{\pm}(x,t) - \theta_{\pm}(0,t)]^2 \rangle=\\
c_{ij} (x,t) \simeq
 \frac{1}{2} \int_0^{\infty} {\rm d} p \ e^{-\alpha^2 p^2} (1-\cos(p x)) \frac{ (-1)^{i+j}}{p^2}\\
 \times  \sqrt{\frac{m_i m_j}{K_i K_j u_i u_j}}    \Big[ \cos((u_i-u_j)pt) - \cos((u_i+u_j)pt) \Big] \ .
\end{multline}
If we neglect the cutoff, the integrals (\ref{ansatz}) can be analytically evaluated. They are of the form
\beq
\begin{array}{ll}
\displaystyle
\int_0^{\infty} {\rm d} p \ (1-\cos( p x)) \cos( u p t)\frac{1}{p^2}
\\ \vspace{-0.2cm} \\
\displaystyle \qquad\qquad=
\begin{cases}
\frac{\pi}{2} (- |t u|+ |x|) &\ \text{if $|x| > |u t|$}
\\ \vspace{-0.2cm} \\
0 & \text{if $|x| < |u t|$}
\end{cases} \ .
\end{array}
\eeq
This shows explicitly the emergence of (sharp) light-cones, associated to each velocity $u$ within the correlation functions. Note that the light-cones are smoothened out (as physically expected) by reintroducing the cutoff.

Correlations like those in Eq.~\eqref{ansatz} appear in the exponent of $C_{\pm}$. Therefore, we expect the approximation \eqref{ansatz} (whose integrand behaves as $ 1/p^2$) to capture only their exponential decay.
A careful analysis should take into account possible power law corrections which come
from the next-to-leading order correction (corresponding to an integrand $\propto 1/p$).
These can be computed explicitly as follows
\beq\label{log_corrections}
\int_{0}^{\infty}\frac{dp}{p}(1-\cos(px))\cos(u pt)=\frac{1}{2}\ln\left|1-\frac{x^{2}}{u^{2}t^{2}}\right|
\eeq
and therefore grow unbounded at large distances.
As we are going to discuss, these terms are actually important, especially for the dynamics of $C_+$:
in fact, there are regimes where the exponential behavior vanishes and power laws become leading.

\subsection{Two-point function: transient, prethermal and stationary state}\label{Sec_two_point}

By looking at the Eq.~(\ref{ansatz}), we can read the leading terms in the two point function \eqref{eq:Cpm}, which presents a very rich behavior.
Without loss of generality, we may assume $u_1>u_2$. Then, we find
\begin{widetext}
\begin{subnumcases}{\label{Eq_two_point}\ln C_{\pm}(x,t) \simeq}
%- \log C_{\pm}(x,t) =
%\begin{cases}
\displaystyle
{-  \frac{\pi}{16}  \frac{m_0 K}{u}  \left[ \frac{1}{K_1^2} 2 u_1 t  +  \frac{1}{K_2^2} 2 u_2 t \mp \frac{2}{K_1K_2} ((u_1+u_2) t  - |u_1-u_2| t)   \right] }
%\qquad
&\text{$x > 2 u_1 t$} \label{Eq_two_point_1}
\\ \vspace{0.2cm}
\displaystyle
{- \frac{\pi}{16}  \frac{m_0 K}{u} \left[ \frac{1}{K_1^2} x +  \frac{1}{K_2^2} 2 u_2 t  \mp \frac{2}{K_1K_2} ((u_1+u_2) t - |u_1-u_2| t)   \right] }
%\qquad
&\text{$2 u_1 t > x  >   (u_1+u_2) t$} \label{Eq_two_point_2}
\\ \vspace{0.2cm}
\displaystyle
 {- \frac{\pi}{16}  \frac{m_0 K}{u} \left[ \frac{1}{K_1^2} x +  \frac{1}{K_2^2} 2 u_2 t  \mp \frac{2}{K_1K_2} (x - |u_1-u_2| t)   \right]}
%\qquad
&\text{$ (u_1+u_2) t > x > 2 u_2 t $} \label{Eq_two_point_3}
\\ \vspace{0.2cm}
\displaystyle
 {- \frac{\pi}{16}  \frac{m_0 K}{u} \left[ \frac{1}{K_1^2} +  \frac{1}{K_2^2}   \mp \frac{2}{K_1K_2}    \right] x
 \mp \frac{\pi}{16}  \frac{m_0 K}{u} \frac{2}{K_1K_2}  |u_1-u_2| t }
%\qquad
&\text{$ 2 u_2 t > x > |u_1-u_2| t $}\label{Eq_two_point_prethermal}
\\ \vspace{0.2cm}
\displaystyle
{-\frac{\pi}{16}  \frac{m_0 K}{u} \left[ \frac{1}{K_1^2} +  \frac{1}{K_2^2}   \right] x }
%\qquad
&\text{$|u_1-u_2| t> x $}\label{Eq_two_point_thermal}
\end{subnumcases}
\end{widetext}
We stress that the expression above only captures the exponential decay of $C_{\pm}$, while power-law corrections are not included. In particular, within this approximation Eq.~(\ref{Eq_two_point_1}) shows no spatial dependence: this does not mean that it does not decay at all, but that the next to leading term should be taken into account. 
By computing it, the corrected expression for $C_+$ in the large distance regime now reads
\begin{multline} \label{eq:C+correct}
C_{+}^{\text{correct}}(x \gg 2u_1 t,t) \propto {|x|^{- \frac{K}{ a \, u} \frac{u_1 u_2}{K_1 K_2}}} \times\\
\times e^{-  \frac{\pi}{16}  \frac{m_0 K}{u} \left[ \frac{1}{K_1^2} 2 u_1 t +  \frac{1}{K_2^2} 2 u_2 t  - \frac{2}{K_1K_2} ((u_1+u_2) t - |u_1-u_2| t)   \right] } \ .
\end{multline}
This result is easy to understand from a physical point of view since
before the quench $\theta_-$ is a massive mode, which does not decay to zero, while
$\theta_+$ is massless with leading power-law correlations. In the regime of large distances and short times therefore
we find the memory of such initial condition.

%\begin{figure}[h!]
%  \begin{center}
%        \includegraphics[height=8.3cm]{densityplots2}
%                 \caption{Logarithm of the correlation functions $C_{+} (x,t)$ (upper row) and $C_{-} (x,t)$ (lower row) after a quench from the ground state of the Hamiltonian (\ref{Semiclassical}) ($T_0=0$), as a function of distance $x$ and time $t$. The exact expressions from Eq.~(\ref{General_corr}) (left panels) are compared with the approximation in Eq.~\eqref{Eq_two_point} (right panels).
%                 The parameters used are $u_1=6$, $u_2=5$, $K_1=20$, $K_2=10$, $g=40$ and $\alpha=1$.
%                 The position of the lightcones derived in Eq.~\eqref{Eq_two_point} are also shown.
%                 The dashed lines correspond to the lightcones at $x= 2 u_1 t,  (u_1+u_2)t, 2u_2 t$ (in the chosen regime $u_1 \approx u_2$ they are close to each other), separating the transient from the prethermal regime.
%                 The dot-dashed line corresponds to the last lightcone at $x= |u_1 -u_2| t$, separating prethermal and final quasi-thermal regime.
%        }\label{fig:densityplot}
%  \end{center}
%\end{figure}

\begin{figure*}[t]
  \begin{center}
        \includegraphics[height=8.5cm]{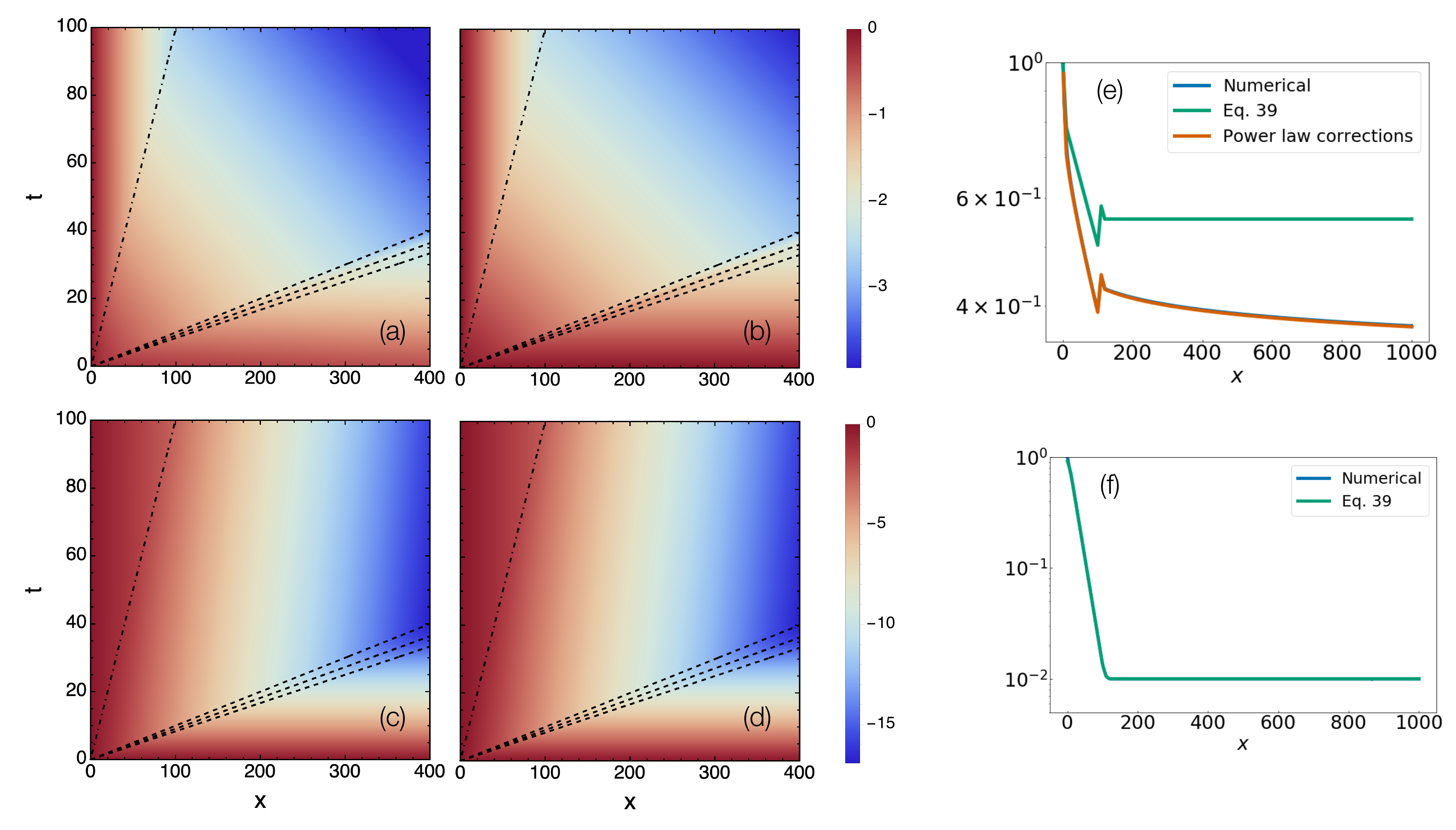}
                 \caption{Logarithm of the correlation functions $C_{+} (x,t)$ (upper row: (a), (b), (e)) and $C_{-} (x,t)$ (lower row: (c), (d), (f)) after a quench from the ground state of the Hamiltonian (\ref{Semiclassical}) ($T_0=0$), as a function of distance $x$ and time $t$. The exact expressions from Eq.~(\ref{General_corr}) (left panels: (a), (c)) are compared with the approximation in Eq.~\eqref{Eq_two_point} (central panels: (b), (d)).
                 The parameters used are $u_1=6$, $u_2=5$, $K_1=20$, $K_2=10$, $g=40$ and $\alpha=1$.
                 The position of the lightcones derived in Eq.~\eqref{Eq_two_point} are also shown.
                 The dashed lines correspond to the lightcones at $x= 2 u_1 t,  (u_1+u_2)t, 2u_2 t$ (in the chosen regime $u_1 \approx u_2$ they are close to each other), separating the transient from the prethermal regime.
                 The dot-dashed line corresponds to the last lightcone at $x= |u_1 -u_2| t$, separating prethermal and final quasi-thermal regime.
                 In the right panels (e),(f) we show a slice of the plots at time $t=10$ comparing the exact numerical results with Eq.~\eqref{Eq_two_point} (with an arbitrary amplitude)
                 and for $C_{+} (x,t)$ the expression corrected by a power law decay in $x$ as in \eqref{eq:C+correct}. 
        }\label{fig:densityplot}
  \end{center}
\end{figure*}

Moreover, from (\ref{Eq_two_point_1}) and its refined version (\ref{eq:C+correct}),
making use of the cluster property, $\lim_{x\to\infty} \langle O(x,t) O(0,t)\rangle  = \langle O(t) \rangle^2 $, we can read the behavior of the one point function
\beq
A_{\pm}(t) \equiv \langle e^{i \sqrt{2} \theta_{\pm}(0,t) } \rangle = e^{- \langle \theta_{\pm}^2(0,t)\rangle } \ .
\eeq
%using the cluster property $\lim_{x\to\infty} C_{\pm}(x,t) = C_{1,\pm}^2(t)$.
For the antisymmetric sector we obtain the exponential decay
\beq
  A_{-}(t) \simeq
e^{- \frac{\pi}{16}   \frac{m_0 K}{u} \left[
 \frac{1}{K_1^2} u_1  + \frac{1}{K_2^2}  u_2
 + \frac{1}{K_1K_2} ((u_1+u_2) - |u_1-u_2|)  \right] t} \ .
\eeq
Note that for the symmetric quench
($K_1=K_2 = K$ and $u_1=u_2=u$) we obtain the results of \cite{Calabrese_QuenchesCorrelationsPRL,Calabrese_QuenchesCorrelationsLong,Bistritzer_IntrinsingDephasingCoupledLL}
with the scaling dimension of $\theta_-$ equal to $h=1/(4 K)$.
Contrarily, for the symmetric sector, due to the power-law correction in \eqref{eq:C+correct}, we find a vanishing one-point function
at all times, i.e.,
$A_{+}(t)= 0$.

From Eq. (\ref{Eq_two_point_thermal}) we see that $C_{\pm}$ reaches a \emph{stationary} state
at short length scales (large times).
Note also that the next to leading order terms in $1/p$, such as integrals of the form (\ref{log_corrections})
do not lead to important time corrections in the limit of large times and that formally
one expects the limit of $t\to\infty$ to be really time independent as all the oscillating factors die out.
 From (\ref{Eq_two_point_thermal}), one can therefore read off an associated correlation length
\beq\label{Corr_length_quench}
\xi_{\text{Q}}^{-1}
%= \frac{\pi}{2} \frac{m_0}{4}  \left( \frac{\cos^2\phi_0}{u_1 K_1} +   \frac{\sin^2\phi_0}{u_2 K_2}    \right)
=\frac{\pi}{16}  \frac{m_0 K}{u} \left( \frac{1}{K_1^2} +   \frac{1}{K_2^2}    \right)
 \ ,
\eeq
equal for both the symmetric and the antisymmetric mode, signaling that the correlations
between the system one and two are lost.
Comparing with (\ref{Eq_corr_length}) this defines an effective temperature
\beq\label{Eq_Teff_stationary}
T^{\text{eff}} = %\frac{m_0}{4}  \frac{ \frac{\cos^2\phi_0}{u_1 K_1} +   \frac{\sin^2\phi_0}{u_2 K_2}  }{
 %\frac{1}{u_1 K_1} +   \frac{1}{u_2 K_2}    }
 \frac{m_0}{4}  \left( \frac{K_1}{K_2} +   \frac{K_2}{K_1} \right) \frac{u_1 u_2}{4 u^2}\ .
\eeq
We dub this regime \emph{quasi-thermal}, to emphasize that in spite of the integrable nature of the system, the final stationary state is well described in the large-scale limit  by a unique correlation length, as in an equilibrium system. We will come back to the precise characterization of such state
 and on the meaning of the ``temperature'' later in the discussions.
Then, if $u_1 \approx u_2$, we have $u_1+u_2 \approx 2 u_1 \approx 2 u_2 \gg |u_1-u_2|$, and
from Eq. (\ref{Eq_two_point_prethermal}), we see that one can define
a quasi-stationary \emph{prethermal state} with correlation length and thus
effective temperature different for the symmetric and the antisymmetric mode
\beq\label{Teff_pm}
T^\text{eff}_{\pm} =
% \frac{m_0}{4}  \frac{\left( \frac{ \cos \phi_0}{\sqrt{u_1 K_1}} \pm   \frac{\sin \phi_0}{\sqrt{u_2 K_2}} \right)^2 }{
 %\frac{1}{u_1 K_1} +   \frac{1}{u_2 K_2}    }
 \frac{m_0}{4} \left( \frac{K_1}{K_2} + \frac{K_2}{K_1} \mp 2  \right) \frac{u_1 u_2}{4 u^2} \ .
\eeq
This is the regime to which the system relaxes in the limit of $u_1=u_2$ and
thus in particular for the symmetric quench.
One can indeed check that for the symmetric quench ($u_1=u_2=u$ and $K_1=K_2=K$)
we recover the results $T^{\text{eff}}_-= {m_0}/{4}$ and $T^{\text{eff}}_+=0$ as
expected from \cite{Calabrese_RevQuenches,Foini_CoupledLLsMassiveMassless} and from the decoupling of the modes.

In Fig.~\ref{fig:densityplot} we show the logarithm of the correlation functions $C_{\pm}(x,t)$ after a quench from the ground state of the Hamiltonian (\ref{Semiclassical}) ($T_0=0$) as a function of distance $x$ and time $t$.
The exact expressions (left panels), numerically computed from Eq.~(\ref{General_corr}), are compared with the small momenta approximation derived in Eq.~\eqref{Eq_two_point} (right panels).
The position of the lightcones are also shown.
 While $C_-$ is well approximated by its exponential decay only in $x$ and $t$ (according to \eqref{Eq_two_point}), for a correct description of $C_+$ power-law correction must be included, 
as is clearly visible by looking at a time slice in Fig.~\ref{fig:densityplot} (right panels).
                 Note that for the parameters chosen, we are in the regime $u_1 \approx u_2$. This is the reason why the regimes in \eqref{Eq_two_point_2}, \eqref{Eq_two_point_3} and \eqref{Eq_two_point_prethermal}, shown as dashed lines in the Figure, are not well separated.
                 Finally, the dot-dashed line corresponds to the last lightcone at $x\approx |u_1 -u_2| t$, separating prethermal and final quasi-thermal regime.

We then focus on the spatial decay of $C_+(x,t)$ and $C_-(x,t)$, for different (fixed) times.
In Fig.~\ref{Fig_CA_T_0} we compare this decay with the equilibrium correlation functions
at temperatures $T^{\rm eff}$ and $T^{\rm eff}_+$ for $C_+(x,t)$ and $T^{\rm eff}$ and $T^{\rm eff}_-$ for $C_-(x,t)$,
which capture the first two exponentially decaying regimes.
In fact, for both correlations the longest time (short distance) shows the crossover between the fully stationary and the prethermal regime at distances around $|u_1-u_2|t$.
After this decay $C_+(x,t)$ is characterized by a non monotonic behavior in the intermediate regimes.
At large distances, differently as compared with $C_-(x,t)$,
it does not saturate but it slowly decreases due to the power law corrections.
The shortest times (long distance) of $C_-(x,t)$ instead show a light-cone like behavior toward a constant value for large
distances.
For the choice of parameters in the figure, the first three light cones in (\ref{Eq_two_point}) are not
separately visible in $C_-(x,t)$ because they are very close (in $C_+(x,t)$ they correspond to the non monotonic behavior).
The presence of the cut-off in  (\ref{General_corr}) also tends to smear out
the sharp transitions in (\ref{Eq_two_point}), as anticipated.

In Fig.~\ref{Fig_CA_CS_T_0} we compare the spatial decay
of $C_-(x,t)$ and $C_+(x,t)$ in the quasi-thermal and in the prethermal regime,
comparing also with the thermal correlations at temperature
$T^{\rm eff}$, $T^{\rm eff}_-$ and $T^{\rm eff}_+$.
The plot shows that the correlation length of the two quantities
coincides in the first (quasi-thermal) regime and is also
compatible with the equilibrium decay at temperature $T^{\rm eff}$.
From this analysis therefore, the last regime can be thought (at the leading order) as a thermal regime, at least for the observables considered here.
However, as we discuss in section \ref{Sec_two_T} the more
robust thermodynamic interpretation of the stationary state
is in terms of two temperatures, one for the first system and one for the second,
which combined give rise coherently to Eq. (\ref{Eq_Teff_stationary}).
At larger distances the two correlations $C_{\pm}$ depart from the thermal
regime and from each other, and agree with an equilibrium-like
behavior at temperature $T^{\rm eff}_-$ and $T^{\rm eff}_+$, respectively (here we explicitly see a dependence on the observable chosen).

\begin{figure}
  \begin{center}
   \includegraphics[height=6cm]{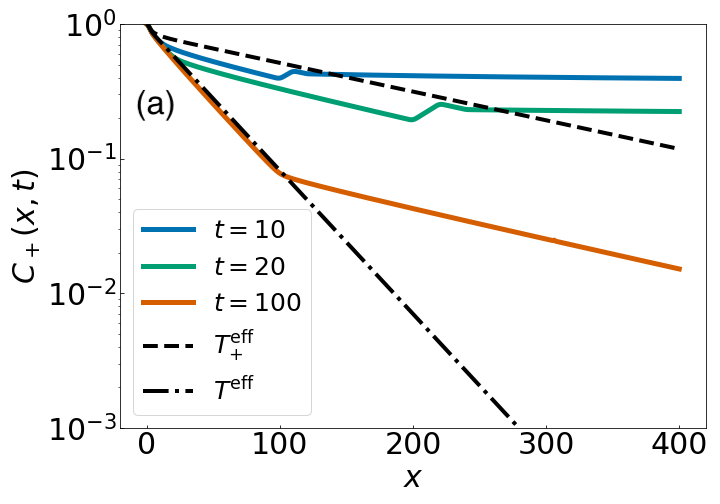}
        \includegraphics[height=6cm]{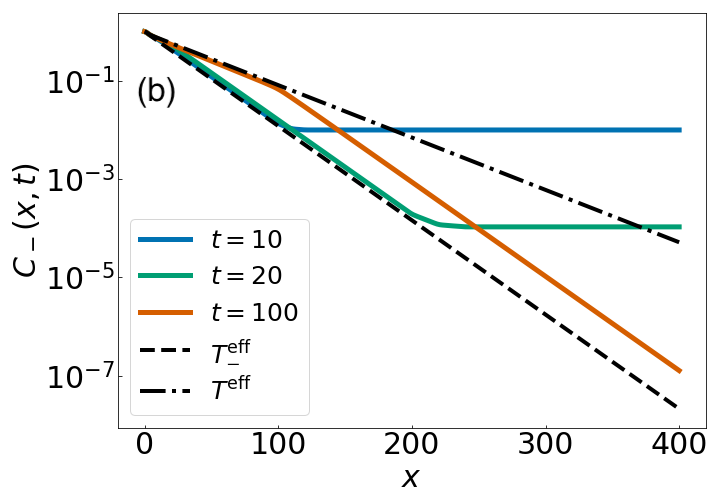}
                 \caption{Space decay of $C_+(x,t)$ (panel (a)) and $C_-(x,t)$ (panel (b)) from (\ref{General_corr})
                 after a quench from the ground state of the Hamiltonian (\ref{Semiclassical}) ($T_0=0$),
%                 with $T_0=0$, namely starting  from the ground state of the Hamiltonian (\ref{Semiclassical}),
                 at different times $t=10, 20, 100$, compared with the equilibrium correlations at temperature
                 $T^{\rm{eff}}$
                 and $T^{\rm eff}_+$ for $C_+(x,t)$ and
                 $T^{\rm{eff}}$
                 and $T^{\rm eff}_-$ for $C_-(x,t)$.
                 The parameters used are $u_1=6$, $u_2=5$, $K_1=20$, $K_2=10$, $g=40$ and $\alpha=1$.
        }\label{Fig_CA_T_0}
  \end{center}
\end{figure}

\begin{figure}
  \begin{center}
        \includegraphics[height=6cm]{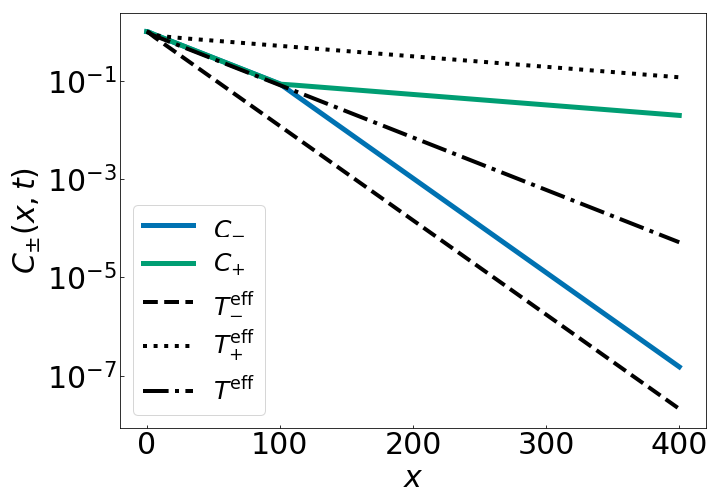}
                 \caption{Space decay of $C_-(x,t)$ and $C_+(x,t)$ from (\ref{General_corr})
                 after a quench from the ground state of the Hamiltonian (\ref{Semiclassical}) ($T_0=0$),
%                 with $T_0=0$, namely starting  from the ground state of the Hamiltonian (\ref{Semiclassical}),
                 at time $t=100$, compared with the equilibrium correlations at temperature $T^{\rm{eff}}$, $T^{\rm eff}_-$
                 and $T^{\rm eff}_+$.
                %The parameters values are the same of Fig.~\ref{fig:densityplot}.
                 The parameters used are $u_1=6$, $u_2=5$, $K_1=20$, $K_2=10$, $g=40$ and $\alpha=1$.
        }\label{Fig_CA_CS_T_0}
  \end{center}
\end{figure}

\subsection{Interpretation as a two temperature system}\label{Sec_two_T}

The final Hamiltonian (\ref{H_final}) has clearly two extensive and different
conserved quantities: the energy of each subsystem.
Therefore we expect to be able to define an effective temperature
associated to each of them from the expectation values of the energy densities of the modes
$\langle \epsilon_{i,p} \rangle \equiv  u_i |p|  \langle b_{i,p}^{\dag} b_{i,p} \rangle $ with $i=1,2$ separately.
In the limit of small momenta the expectation of each mode is dominated by a constant
term equal for all the modes.
Thanks to a classical equipartition approximation this allows us to interpret such constant
 as the effective temperature
of the two systems \cite{Foini_CoupledLLSchmiedtmayerMassive}.
In particular we have
\beq\label{T1_T2}
\begin{array}{l}
\displaystyle
\langle \epsilon_{1,p} \rangle \simeq
%\frac{m_0}{4}  \cos\phi_0^2 =
\frac{m_1}{4} = \frac{m_0}{8}  \frac{u_1}{K_1} \frac{K}{u}  \equiv T_1^\text{eff} ,
\\ \vspace{-0.2cm} \\
\displaystyle
 \langle \epsilon_{2,p} \rangle \simeq
%\frac{m_0}{4} \sin\phi_0^2   =
 \frac{m_2}{4} = \frac{m_0}{8}  \frac{u_2}{K_2} \frac{K}{u} \equiv T_2^\text{eff} .
\end{array}
\eeq
Note that
\beq
\langle \epsilon_{1,p} \rangle + \langle \epsilon_{2,p} \rangle = \frac{m_0}{4} \ ,
\eeq
as for the symmetric quench \cite{Calabrese_RevQuenches,Foini_CoupledLLsMassiveMassless}. However, contrary to the symmetric limit where all the energy is stored in the antisymmetric sector (being isolated from the symmetric one), in the general case part of it is shared with the symmetric mode as well.

Note that if one supposes the two systems equilibrated at different temperatures,
the correlation length associated to the decay of $C_{\pm}^{\text{eq}}$ turns out to be
\beq\label{Eq_corr_length2}
\xi_{T_1,T_2}^{-1} = \frac{\pi}{2} \left( \frac{T_1}{u_1 K_1}+\frac{T_2}{u_2 K_2} \right) \ ,
\eeq
thus generalizing the expression (\ref{Eq_corr_length}).
This is perfectly consistent with the effective temperatures (\ref{T1_T2}) and
the post quench correlation length (\ref{Corr_length_quench}). Specifically, the \emph{unique} effective temperature, which we can read off from $C_{\pm}$ at large times, is related to the ones in  (\ref{T1_T2}) through
\beq \label{eq:Teff}
T^{\text{eff}}= \left( \frac{T_1^{\text{eff}}}{u_1 K_1} +   \frac{T_2^{\text{eff}}}{u_2 K_2}  \right)/  \left(  \frac{1}{u_1 K_1} +   \frac{1}{u_2 K_2}  \right) \ .
\eeq
The two-temperature interpretation is further sustained by an FDT (\emph{fluctuation-dissipation theorem}~\cite{Callen_FDT,Chou_ReviewFDT,Cugliandolo_OutOfEquilibriumFDT,Bouchaud_OutOfEquilibriumGlassySystems}) argument, analyzing the correlation and the response functions associated to
 the Green's functions of the two systems -- in the limit of small $\omega$ and small $p$ (see Appendix \ref{Appendix_FDT}).
Note however that, while this interpretation is definitely more robust, 
it still is an approximation of the true underlying generalized Gibbs ensemble (GGE)\cite{Rigol_RelaxationHardCoreBosons, Rigol_GGE}. Moreover, this analysis requires more attention when trying to generalise to all observables (see the discussion section).

We close this section with an interesting remark. Knowing that in our quench (when $u_1\neq u_2$)
the correlations between the two systems are lost in the stationary regime, one can expect the quasi-thermal state to which the system evolves to coincide with the final state reached by the same system of two bosons but after two independent quenches with initial energies (or initial masses, equivalenty) fixed by \eqref{T1_T2}.  As main difference, in this simpler quench, correlations are absent also in the initial state.
If fact, since each $H_i$ ($i=1,2$) describes a conformally invariant system, one can directly apply the results of \cite{Calabrese_QuenchesCorrelationsPRL,Calabrese_QuenchesCorrelationsLong} for the correlation functions to see that at largest times
\begin{multline} \label{eq:corrCFT}
%C_{\pm} (x,t) =
 \langle e^{i \sqrt{2} (\theta_{\pm} (x, t) -\theta_{\pm} (0, t)) } \rangle_{m_1, m_2}= \\
 =\langle e^{i (\theta_1 (x, t) -\theta_1 (0, t)) } \rangle_{m_1}  \langle e^{\pm i (\theta_2 (x, t) -\theta_2 (0, t)) } \rangle_{m_2}\\\simeq e^{-\frac{\pi}{2} \left(\frac{h_{1} m_1}{u_{1}}+\frac{h_{2}m_2}{ u_{2}}\right)x} =e^{-\frac{\pi}{8}\left(\frac{m_{1}}{u_{1}K_{1}}+\frac{m_{2}}{u_{2}K_{2}}\right)x} \ .
\end{multline}
The expectation value $\langle \cdot \rangle_{m_1,m_2}$ in the first line is taken on a \emph{factorized} state characterized by mass $m_i$ for the $i$-th system, which, therefore, simply splits in expectation values over the two systems (second line). The results of \cite{Calabrese_QuenchesCorrelationsPRL,Calabrese_QuenchesCorrelationsLong} have been applied to each $\langle \cdot\rangle_{m_i}$.
In the last step, we used the explicit form $h_i=1/(4K_i)$ for conformal dimensions of the (primary) operators $e^{\pm i \theta_i (x, t)}$~\cite{yellowbook}.
We stress, however, that, even if the result \eqref{eq:corrCFT} is consistent with the last regime with associated effective temperature \eqref{eq:Teff}, the transient and prethermal regime are not captured by this simple picture.

\section{Quench from a thermal state: corrections due to the initial temperature}\label{Sec_thermal_state}

If the initial state is prepared at finite temperature $T_0$, the full expression for the correlation function is still the one in Eq.~(\ref{General_corr}).
Now, however, one sees that, differently from the quench from the ground state, the leading contribution as $p\to 0$ includes a term coming from the massless mode.
One can in principle carry a similar analysis as the one of the section~\ref{sec:GS} (notice in particular that Eq.~\eqref{Eigenmodes_dynamics} remains true also when starting from a thermal state), leading to different regimes during the evolution. In particular in Appendix \ref{Appendix_thermal_quench} we sketch
the derivation of the leading order term contributing to $C_{\pm}(x,t,T_0)$ showing that the same light cones as for $T_0=0$ appear, with different correlation lengths and coherence times.
Here however, we focus on the last two regimes (at large times), being the most relevant for the relaxation dynamics.
As before, indeed, they allow for a definition of a prethermal and a quasi-thermal correlation length,
for both the symmetric and the antisymmetric mode.
The associated prethermal effective temperatures now read
\beq\label{Teffpm_T0}
\begin{array}{l}
\displaystyle
T^{\text{eff}}_{\pm} =
\frac{m_0}{4} \left( \frac{K_1}{K_2} + \frac{K_2}{K_1} \mp 2  \right) \frac{u_1 u_2}{4 u^2}
\text{cotanh}\left( \frac{m_0}{2 T_0}\right)
\\ \vspace{-0.2cm} \\
\displaystyle
\qquad + \Big(   (1 \pm 1)  \frac{ u_1 u_2 K_1 K_2}{4 u^2 K^2}   +  \frac{1}{8} \frac{ (u_1 \pm  u_2)^2}{u^2}   \Big) T_0 \ .
\end{array}
\eeq
For the symmetric quench we recover $T^{\text{eff}}_- = \frac{m_0}{4}\text{cotanh}\left( \frac{m_0}{2 T_0}\right) $
as in \cite{Sotiriadis_ThermalQuench} and $T^{\text{eff}}_+=T_0$, as expected from the decoupling of the modes.
A crucial observation here is that in this symmetric limit, the antisymmetric sector is almost unaffected by the true temperature of the system: in fact $T^{\text{eff}}_- \simeq {m_0}/{4}$, namely it is independent of $T_0$ (as long as it is low), while the thermal fluctuations are present only in the symmetric mode, as reflected by its effective temperature. The reason is that, while $\theta_1$ and $\theta_2$ are subject to thermal fluctuations, those cancel out in their difference (namely in $\theta_-$), while remaining present in their sum (i.e., in $\theta_+$)~\cite{Langen_PrethermalizationNearIntegrable}.
Importantly, this picture completely changes as soon as an asymmetry is induced in the parameters $u_i,K_i$ associated with the two tubes. In fact, Eq.~\eqref{Teffpm_T0} clearly shows a correction linear in $T_0$ for the effective temperature. To be more precise, for such linear correction to be present in the antisymmetric mode as well, different sound velocities, i.e., $u_1 \neq u_2$, are needed (while a difference in the Luttinger parameters $K_i$ does not seem to play a main role here).
In this case, the initial temperature plays a crucial role in the decay of all correlation functions. Specifically, since the term proportional to $ T_0$ in \eqref{Teffpm_T0} is always positive, it leads to a \emph{faster} decay of $C_{\pm}$.

The final regime is instead described by
\begin{multline}\label{Teff_T0}
T^{\text{eff}} =
\frac{m_0}{4} \left( \frac{K_1}{K_2} + \frac{K_2}{K_1}   \right) \frac{u_1 u_2}{4 u^2}
\text{cotanh}\left( \frac{m_0}{2 T_0}\right) \\
\qquad+ \Big(  \frac{ u_1 u_2 K_1 K_2}{4 u^2 K^2}   +  \frac{1}{8} \frac{ u_1^2 +  u_2^2}{u^2}  \Big) T_0 \ .
\end{multline}
which also shows a term depending linearly on the initial temperature, leading to faster decaying correlations.

\begin{figure}
  \begin{center}
        \includegraphics[height=6cm]{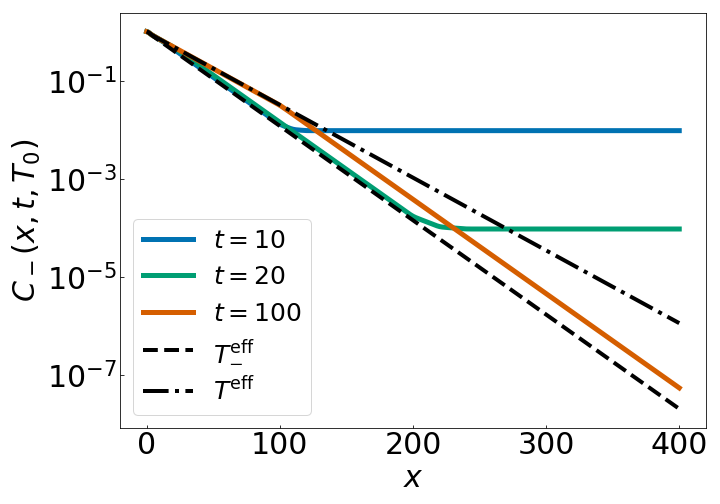}
                 \caption{Space decay of $C_-(x,t,T_0)$ from (\ref{General_corr}) with
                  $T_0=0.5$,
                 at different times $t=10, 20, 100$, compared with the equilibrium correlations at temperature $T^{\rm{eff}}$
                 and $T^{\rm eff}_-$.
                 The parameters used are $u_1=6$, $u_2=5$, $K_1=20$, $K_2=10$, $g=40$ and $\alpha=1$.
        }\label{Fig_CA_beta_2}
  \end{center}
\end{figure}

\begin{figure}
  \begin{center}
        \includegraphics[height=6cm]{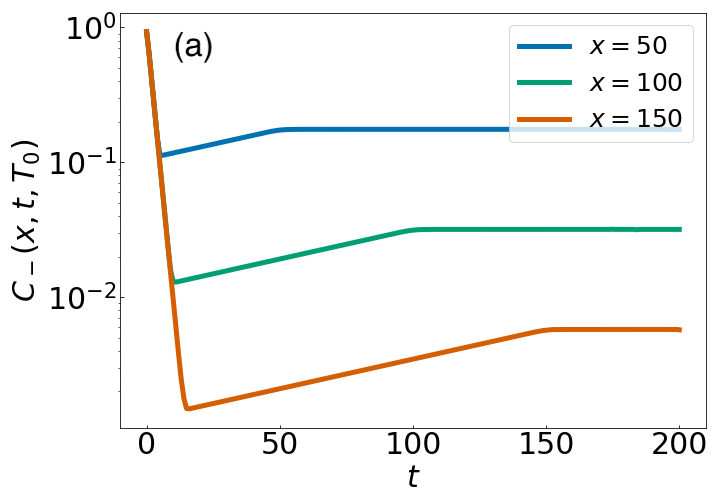}
         \includegraphics[height=6cm]{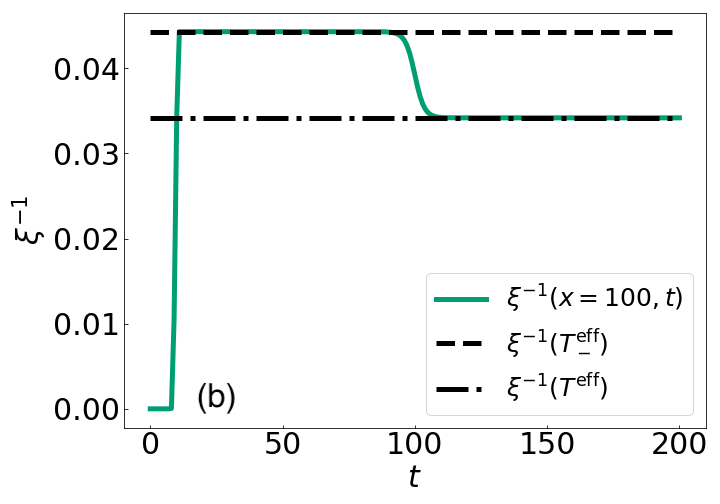}
                 \caption{Panel (a): Time dependence of $C_-(x,t)$ from (\ref{General_corr}) with
                  $T_0=0.5$,
                 at different points in space $x=50, 100, 150$.
                 Panel (b): Inverse correlation length (obtained as the spatial derivative of
                 the exponent of Eq. (\ref{General_corr})) at distance $x=100$ as a function of the time,
                 compared with the equilibrium correlation length at temperature $T^{\rm{eff}}$
                 and $T^{\rm eff}_-$.
                 The parameters used are $u_1=6$, $u_2=5$, $K_1=20$, $K_2=10$, $g=40$ and $\alpha=1$.
        }\label{Fig_CA_time_beta_2}
  \end{center}
\end{figure}

We mention that the limit of shallow quench $m_0\to0$ (which amounts to
doing nothing to the system) does not reproduce
the equilibrium result $T^{\text{eff}} = T_0$.
This signals that the limit of small momenta $p\to0$ used in deriving $T^{\text{eff}}$ in \eqref{Teff_T0} does not commute with the limit $m_0\to 0$.

In Fig.~\ref{Fig_CA_beta_2} we show how the effective temperatures derived above capture the main decay of the correlation functions, also in this quench starting from a thermal state.
In particular, it shows the spatial decay of $C_-(x,t)$ starting
from a thermal state at temperature $T_0=0.5$ and different times.
The correlation lengths in the quasi-thermal and in the prethermal regime are compared
with the one at equilibrium at temperature $T_-^{\rm eff}$ from (\ref{Teffpm_T0})
and $T^{\rm eff}$ from (\ref{Teff_T0}). The plateau attained at large distances (short times) is instead a property of the (massive) initial condition.

In Fig.~\ref{Fig_CA_time_beta_2} we plot the same correlator as a function of time.
The top panel shows the time dependence of $C_-(x,t)$ again starting
from a thermal state at temperature $T_0=0.5$ and different points in space.
In the bottom panel, instead, the inverse correlation length
at fixed distance is shown, as obtained from the spatial derivative of the exponent in Eq.~(\ref{General_corr}).
We see that in the regime $u_1,u_2 \gg |u_1-u_2|$ there is an intermediate prethermal
regime where the correlation length is compatible with the equilibrium one at temperature
$T^{\rm eff}_-$. At later times this quantity crosses over towards the asymptotic
regime, compatible with the equilibrium one at temperature $T^{\rm eff}$.

\section{Discussions} \label{sec:discussion}

Let us make some comments about the results obtained
in the previous sections, also in comparison with previous works.

To start with, since in our analysis we considered the generic case of different $u_i$ and $K_i$ (i=1,2),
it is worth stressing the different role that these two parameters play
in the dynamics.
In fact, while we saw that, in the Hamiltonian~\eqref{H_final}, which governs the evolution after the quench, a coupling between symmetric and antisymmetric sector is present as soon as
the systems one and two differ in either of the parameters (cfr. Eq.~\eqref{H_piu_meno}),
the consequences of having different $K_i$ or different $u_i$
separately are not the same.
If $u_1=u_2$, then the correlation functions (\ref{Eq_two_point})
are much simplified and only one lightcone appears, with the dynamics \emph{never} reaching the final regime (\ref{Eq_two_point_thermal}). This means that in this case symmetric and antisymmetric sectors show different effective final temperatures.
Moreover the linear correction of the effective temperature of the antisymmetric sector (\ref{Teffpm_T0})
 due to the initial temperature $T_0$  vanishes.
Such observations further suggest that a sort of decoupling between different sectors still exists.
In fact, in the final Hamiltonian (\ref{H_final}), one could rescale the field
$\theta_i$ by $\sqrt{K_i}$  and $n_i$ by $1/\sqrt{K_i}$ in such a way to respect canonical
commutation relations and end up in a system of effectively identical TLLs,
allowing for additional conservation laws than those associated to $H_1$ and $H_2$
(as for the symmetric quench).
On the other hand having different $u_i$ but same $K_i$ does not
modify the generic (richer) picture outlined in (\ref{Eq_two_point}), which is characterized
by the presence of multiple lightcones and regimes.
And, in fact, in this second
case, the difference in the two tubes can not be reabsorbed in a rescaling
of the variables similar to the one above.
One peculiarity of this limit, however, is the fact that the effective
temperature of the symmetric mode in the prethermal regime (\ref{Teff_pm}) is zero.

Let us now turn to the final stationary regime reached by the dynamics.
As we discuss in section \ref{Sec_two_T} and Appendix \ref{Appendix_FDT}, such regime for many observables and in an RG sense (namely at large scales), is compatible with an equilibrium-like
result associated to the two systems thermalized at temperatures $T_1^{\rm eff}$ and $T_2^{\rm eff}$,
in accordance with the classical equipartition theorem and the FDT in its classical
(low frequency) approximation. While this appears as  a generalization to a two
temperature equilibrium  state of previous results \cite{Calabrese_QuenchesCorrelationsPRL,Gring_Prethermalization,Foini_CoupledLLsMassiveMassless}, it might sound surprising
given that the underlying dynamics conserves the energy of each mode.
In fact, a GGE \cite{Rigol_RelaxationHardCoreBosons, Rigol_GGE} would rather appear from Eq. (\ref{General_corr}),
if we would take into account the full dependence on the momenta $p$ in the integrals.
However, as we have seen, the dynamics of the vertex operators of the antisymmetric mode (see the bottom panels of Figure \ref{fig:densityplot})
and the stationary part of the symmetric one (see the regime within the first lightcone at short distances in the top panels of Figure \ref{fig:densityplot})
are well captured by the leading order term in $p\to 0$ of the integrands which gives the expressions (\ref{Eq_two_point})
and in particular (\ref{Eq_two_point_thermal}).
This fact by itself is quite remarkable since this is not usually the case for quenches in the LL (see \cite{Cazalilla_MasslessQuenchLL} as main reference), where the underlying GGE describing the steady state is not thermal at all. 
Given a GGE of the form $\rho_{GGE}=\exp {(-\sum_p \beta(p) n(p))}$ (with $n(p)$ the conserved charges and $\beta(p)$ the associated temperatures, labelled by momentum $p$), this might or not be well approximated by a thermal ensemble, depending on the behaviour of the (inverse) temperatures $\beta(p)$ as a function of $p$. 
In particular, if we focus on the large scale limit, the modes that matter are the low energy ones and their behaviour is indeed what makes our quench in the tunnelling strength very different from the one in the interaction studied in Ref.~\cite{Cazalilla_MasslessQuenchLL}.
Nonetheless, an example where the non-thermal behaviour clearly emerges also in our setup is given by the density-density correlations. In this case the leading term at large scale seems related to the first \emph{singularity} in the small $p$ expansion (rather than a simple small $p$ expansion), giving rise to a power law decay. From the GGE point of view this means that the first term in the small $p$ expansion of $\beta(p)$ is not enough to capture the leading behavior. Physically, this contribution can be traced back to the presence of the massless mode, that now becomes leading.
A complete analysis of this kind of correlations will be given elsewhere, from a different perspective~\cite{Ruggiero_QuenchCoupledCFTs}.
%Moreover, while stationary expectation values of several observables are in agreement with this interpretation, one can also find counter-examples of quantities behaving in a very non-thermal way (e.g., the density-density correlations). However in such case a more careful treatment is needed, which is beyond the scope of this paper and will be analized elsewhere~\cite{Ruggiero_QuenchCoupledCFTs}.

%The accuracy of the approximation made in (\ref{Eq_two_point})
%actually, is quite good also to describe the prethermal phase of the symmetric mode (as can be seen in the
%corresponding regime of Figure \ref{fig:densityplot} and in Figure \ref{Fig_CA_CS_T_0}).

Moreover, in our discussion we referred to the regime (\ref{Eq_two_point_prethermal}) (at least in the limit of $u_1\simeq u_2$) as a \emph{prethermal} one, in analogy
with the work \cite{Gring_Prethermalization} (note however that, given the relaxation to a GGE discuss above, rather than a true thermalization, the term \emph{pre-relaxation}, which already appeared in literature~\cite{Fagotti_Prerelaxation_Free,Bertini_Prerelaxation_Interacting}, would be more appropriate).
%(there, the dynamics
%stops at this prethermal regime due to the decoupling between
%symmetric and antisymmetric sectors, which holds
%in the symmetric setting).
More generally, prethermalization has been discussed in many works and it is often
associated to a slow evolving intermediate state attained by the system before a complete relaxation takes place,
as it happens in integrable systems in presence of a small integrability
breaking perturbation \cite{Bertini_LightconesIntegrabilityBreaking,Bertini_PrethermalizationIntegrabilityBreaking,Langen_PrethermalizationNearIntegrable,Marcuzzi_PrethermalizationNonIntegrableSpinChain,Kollar_PrethermalizationIntegrableSystems, Mitra_IntegrabilityBreakingOutOfEquilibrium} or in other more exotic scenarios as in~\cite{Alba_NewPrethermalizationMechanism}.
In this sense, the fact of considering different $u_i$ can be seen as a
symmetry breaking mechanism that
removes the degeneracy of the hamiltonian driving the dynamics.
In fact, from Figure \ref{Fig_CA_time_beta_2} (particularly if focusing on the inverse correlation length, bottom panel) one clearly sees the presence of a first rapid transient regime, followed by a quasi-stationary one for a relatively large time (divergent in the limit $u_1 \to u_2$) and later evolving towards its asymptotic value. Note however that in order for the final state to be reached, the prethermal plateau cannot be really time-independent and this is in fact clearly visible when looking at the correlation function $C_-$ itself (the top panel of the same Figure), which shows a slow ramp towards the final stationary regime.
Note that this ramp can be increasing or decreasing according to
the sign of $T_-^{\rm eff}-T^{\rm eff}$, which can be tuned upon varying $T_0$.

About the main experimental implications of our results,
one of the most surprising effects of considering two TLLs with different parameters is the
(positive) \emph{linear correction} in $T_0$ to the effective temperature $T_{-}^{\rm eff}$ of the antisymmetric sector, in contrast to the insensitivity of the same in the symmetric scenario~\cite{Gring_Prethermalization}.
This implies faster decaying correlations and it might be a non negligible effect in the dynamics, given the relative high temperature at which experiments are carried out.
For example, we would expect a similar correction to take place in the experiment discussed in~\cite{Pigneur_RelaxationPhaseLockedState}: there, in principle, the very same analysis can be carried out, while for now a theoretical understanding of the observed ``effective'' dissipation mechanism is still missing~\cite{vanNieuwkerk_QuenchSineGordonSelfConsistentHarmonicApprox,vanNieuwkerk_TunnelCoupledBoseGasesLowEnergy,Pigneur_EffectiveDissipativeModel} (see also~\cite{Polo_Josephson_damping_head-to-tail} where the same problem is studied but within a different geometrical setup).

Remarkably, the phenomenological description of the unbalanced splitting
protocol of Ref.~\cite{Langen_UnequalLL} for two bosonic tubes at different densities agrees in many aspects with the overall picture emerging from our general analysis of the quench dynamics in unbalanced TLLs coupled by tunnelling. 
There, in particular, the transition from a prethermal to a thermal regime, both characterized by an exponential
decay of correlation functions, with a multi light-cone dynamics signalling the sharp transition
between different correlation lengths was found, as well as
an additive correction proportional to the initial true temperature $T_0$ to the final effective temperature, shared
by both the symmetric and the antisymmetric sector.
Such effects are indeed a consequence of the form of the initial correlations (fixed by phenomenological reasoning in Ref.~\cite{Langen_UnequalLL}, while derived in our case),
whose leading term behaves as $p^{-2}$ (see Eqs. (\ref{Eq_corr_theta_T0}) and (\ref{Eq_corr_n_T0})). 
%This behavior leads to the exponential decay and a correction in the initial temperature showing up in the final correlation length.
Moreover, the relation $T^{\text{eff}}=(T^{\text{eff}}_- +T^{\text{eff}}_+) /2$ in \cite{Langen_UnequalLL}, connecting the
prethermal and the thermal effective temperatures, is found to hold in our more general setting (cfr. Eqs.~(\ref{Teffpm_T0}) and (\ref{Teff_T0})).

There are however some interesting differences. 
In particular the case of density imbalance, $\rho_1 \neq \rho_2$, studied in Ref.~\cite{Langen_UnequalLL}, leads to the vanishing of the $n_{\pm}$ mixing term in \eqref{H_piu_meno} (i.e., $\Lambda_n=0$ in \eqref{lambdas}), 
while more general imbalances (coming, for example, from different 1d interactions $U_i$) would allow for the presence of such a term. 
Moreover, due to the difference in the two protocols (namely, the starting point of \cite{Langen_UnequalLL} is an imbalanced splitting of a single tube, while we start directly from two different tubes with non-zero tunnelling), our temperatures show a
different dependence on the density as one can easily check by substituting the parameters (\ref{micro_parameters}) in our expressions.
In particular, note that, in our protocol, if the imbalance is just in the densities and in the limit $T_0 \to 0$, we get
that the temperatures of the two systems given in (\ref{T1_T2})
are the same, i.e., $T_1^{\text{eff}}=T_2^{\text{eff}}$, and therefore they also coincide with the final temperature of the symmetric and antisymmetric sectors. This, however, is not the case anymore at finite temperature $T_0 \neq 0$, 
and a \emph{linear correction} in $T_0$ appears also to the prethermal temperature $T^{\text{eff}}_-$ of the \emph{asymmetric} sector, due to the difference between the two velocities. 

Some of the effects mentioned above were also analyzed in Ref.~\cite{Kitagawa_DynamicsPrethermalizationQuantumNoise}, 
which was considering a two ``spin" mixture, analagous to our two leg ladder system, and thus discuss a quench for a similar Hamiltonian.
However in their case the initial state is chosen to be a \emph{factorized} state of the symmetric and antisymmetric parts.
For our quench this is not the case and the initial state does not simply factorize, thus leading to different time evolutions.

It would be very interesting to test the previously highlighted features, displaying strong differences as compared to the equal TLLs scenario.
This could be done e.g. in experiments similar to the ones of the Vienna's group~\cite{Kuhnert_ExpEmergenceCharacteristicLength1d,Langen_ThermalCorrelationsIsolatedSystems, Langen_ExperimentGGE,Gring_Prethermalization}. Given the importance played by the sound velocities in the dynamics, the presence of the harmonic confinement potential (where the gas is trapped) leading to a spatially dependent velocity is 
clearly a highly unwanted complication. Fortunately, however, the recent realization of boxlike potentials in such experiments~\cite{Jorg_privatecommunication} shows great promise that the features analyzed in the present paper could be tested in a near future. 
Note that although here we mainly focused on vertex correlators, our analysis gives a full diagonalization of the problem, so in principle other correlation functions are also easily accessible.

\section{Conclusions}\label{sec:conclusions}

In this work we have studied a quench in the tunnelling strength
of two TLLs with different parameters, under a quadratic approximation for the initial tunnel coupling term. 

Our results show that the fact of considering two unqual
systems leads to a much richer physics than the one observed in the symmetric scenario.
This is manifested, for instance, in the emergence of multiple light cones.
Moreover, under this dynamics, the prethermal regime discussed in \cite{Gring_Prethermalization}
is followed by a final stationary state, that we dubbed \emph{quasi-thermal}, where symmetric
and antisymmetric sectors display the same effective temperature (spatial decay).
Due to the coupling between the symmetric/antisymmetric sectors, one observes also an important
effect of the initial temperature on the correlation length (effective temperature) measured
via the decay of the antisymmetric mode, which otherwise would be
only slightly modified in the limit of large initial masses.

Our prediction could be tested in experiments similar to the ones performed~\cite{Kuhnert_ExpEmergenceCharacteristicLength1d,Langen_ThermalCorrelationsIsolatedSystems, Langen_ExperimentGGE,Gring_Prethermalization} for the symmetric quenches.

Beyond the current work the generalized Bogolioubov transformations developed in this paper allow us
to address also different settings and a natural sequel of this work would be to
consider the opposite quench, namely from a massless (uncoupled) initial condition
to a massive (coupled) dynamics~\cite{Ruggiero_MasslessMassiveAsymmLLs}. 
Another interesting direction to pursue is to understand the solution of the dynamics
outlined in this work from the perspectives
of a conformal field theory (CFT) approach~\cite{Ruggiero_QuenchCoupledCFTs}, generalizing the
ideas of~\cite{Calabrese_QuenchesCorrelationsPRL,Calabrese_QuenchesCorrelationsLong,Calabrese_RevQuenches}
to the quench of two independent CFTs coupled by a (conformal) initial condition.

\acknowledgments

We thank J\"org Schmiedmayer and E. Demler for important discussions and for pointing
us Ref.~\cite{Langen_UnequalLL} and the study of the asymmetric quench in Ref. \cite{Kitagawa_DynamicsPrethermalizationQuantumNoise}.
We also thank  Vincenzo Alba and Pasquale Calabrese for useful discussions and J\'er\^ome Dubail for comments on the manuscript.
This work is supported by ``Investissements d'Avenir" LabEx PALM
(ANR-10-LABX-0039-PALM) (EquiDystant project, L. Foini)
 and by the Swiss National Science Foundation under Division II. 

\appendix

\section{Bogoliubov transformation} \label{appendix:Bogoliubov}

We want to diagonalize the hamiltonian \eqref{Semiclassical}. To this aim, we go to Fourier space, where it can be decomposed as
\begin{equation}
H_{\text{initial}}^{\text{SC}} =
\sum_{p\neq 0} \boldsymbol{b}^{\dag}_p H_p \boldsymbol{b}_p,
\end{equation}
with $\boldsymbol{b}^{\dag}_p =(b_{1,p}^{\dag} \, b_{1,-p} \, b_{2,p}^{\dag}  \, b_{2,-p})$.
Above, $H_p$ is of the form
\beq \nonumber
%H_{p}=
\begin{bmatrix}w_{1,p}+D_{1,p} & -D_{1,p} & -C_{p} & C_{p}\\
-D_{1,p} & w_{1,p}+D_{1,p} & C_{p} & -C_{p}\\
-C_{p} & C_{p} & w_{2,p}+D_{2,p} & -D_{2,p}\\
C_{p} & -C_{p} & -D_{2,p} & w_{2,p}+D_{2,p}
\end{bmatrix}
\eeq
and ($i=1,2$)
\beq
w_{i,p}=\frac{u_{i}|p|}{2},\quad D_{i,p}=\frac{g}{8}\frac{1}{K_{i}|p|},\quad C_{p}=\sqrt{D_{1,p}D_{2,p}} \ .
\eeq
The problem is thus reduced to the diagonalization of the $4\times4$ matrix $H_{p}$.
This can be achieved via a Bogoliubov transformation~\cite{Bogoliubov_Bogoliubov,Valatin_Bogoliubov}, which is a linear transformation $B$ on the bosons $\boldsymbol{b}_p$.
Restricting to real transformations, it has $16$ free parameters. However, it has to satisfy some constraints \cite{Elmfors_Bogoliubov}.
First of all, the $4$ bosonic modes defining $\boldsymbol{b}_p$ are not independent, but are instead related (in pairs) by $p\to-p$. This reduces the free parameters to 8, and constrains the corresponding Bogoliubov matrix to be of the form
\begin{equation}
B\equiv\begin{pmatrix}\alpha & \beta\\
\gamma & \delta
\end{pmatrix},\quad\alpha=\begin{pmatrix}\alpha_{1} & \alpha_{2}\\
\alpha_{2} & \alpha_{1}
\end{pmatrix}
\end{equation}
and the same for $\beta,\gamma,\delta$.
Moreover, we want $B$ to preserve canonical commutation relations, i.e.,
\begin{equation}
[\boldsymbol{b}_{p,\mu}, \boldsymbol{b}_{p, \nu}^{\dagger}]=J_{\mu\nu},\quad J\equiv
\begin{bmatrix}1\\
 & -1\\
 &  & 1\\
 &  &  & -1
\end{bmatrix}
\end{equation}
where $\mu, \nu = \{1, 2, 3, 4 \}$.
This requirement leads to the condition
\beq \label{symplectic}
B JB^{\text{t}} {=}J \ ,
\eeq
namely $B$ must be a symplectic matrix.
Eq.~\eqref{symplectic} is equivalent to
\begin{eqnarray*}
(\alpha_{1}^{2}-\alpha_{2}^{2})+(\beta_{1}^{2}-\beta_{2}^{2})	=	1\\
(\gamma_{1}^{2}-\gamma_{2}^{2})+(\delta_{1}^{2}-\delta_{2}^{2})	=	1\\
\alpha_{1}\gamma_{1}-\alpha_{2}\gamma_{2}+\beta_{1}\delta_{1}-\beta_{2}\delta_{2}	=	0\\
\alpha_{1}\gamma_{2}-\alpha_{2}\gamma_{1}+\beta_{1}\delta_{2}-\beta_{2}\delta_{1}	=	0 .
\end{eqnarray*}
If we take $(\alpha_{1}^{2}-\alpha_{2}^{2})\geq0$ and the same for $\beta,\gamma,\delta$, the solutions can be parametrized by a Bogoliubov matrix of the form given in Eq.~\eqref{Bogoliubov_matrix}, with $B\equiv B(\hat\varphi_p)$ depending on a set of $4$ parameters
 $\hat\varphi_p=\{\varphi_{1,p},\varphi_{2,p},\Delta_p,\phi_p\}$.
Finally, their value is uniquely fixed by the requirement for $B$ to diagonalize $H_p$.
Note that this is not a standard diagonalization problem, because of the symplectic nature of $B$.
The standard procedure \cite{Tsallis_Bogoliubov,vanHemmen_Bogoliubov} amounts to finding the spectrum of $H_p$, by introducing the matrix $H_p J$. This one can now be diagonalized in a standard way, meaning via a unitary transformation $T$ as
$$T^{-1}(H_pJ)T=\Lambda_{p}J \ , $$
with $\Lambda_{p}$ diagonal (the corresponding spectrum in our case is given by Eq.~\eqref{Eigenvalues} in the main text). Eventually, one imposes $B^{\text{t}}H_pB=\Lambda_p$. This fixes the parameters $\hat\varphi_p$ to be of the form given in Eq.~\eqref{parameters}.

\section{Derivation of $C_{\pm}(x,t,T_0)$, Eq.~\eqref{General_corr}} \label{sec:Wpm}

We start by considering the logarithm of $C_{\pm}(x,t,T_0)$ defined in \eqref{eq:Cpm}, i.e.,
%
%\begin{multline}
\beq
\label{eq:logCpm}
\langle[\theta_{\pm}(x,t)-\theta_{\pm}(0,t)]^{2}\rangle_{T_0}.
\eeq
%\end{multline}
%
If we expand the square inside the expectation value, it is the sum of 4 terms of the form
\beq \label{thetapmxy}
\langle\theta_{\pm}(x,t)\theta_{\pm}(y,t)\rangle_{T_0}=\frac{1}{2} \sum_{i,j=1,2} (\pm 1)^{i+j} \langle\theta_{i}(x,t)\theta_{j}(y,t)\rangle_{T_0}.
\eeq
The problem is thus reduced to the evaluation of correlation functions of $\theta_i (x,t)$ ($i=1, 2$).
This can be achieved, as in section~\ref{sec:eigdynamics}, by looking at the the dynamics of $\theta_{i}(p,t)$.
An alternative way, however, it to use the expansion of the fields $\theta_i (x,t)$ in terms of the creation/annihilation operators $b_{p,i}(t)$ (at $t=0$ it is given by Eq.~\eqref{expansion_theta} in the main text, with $b_{i,p}\equiv b_{i,p} (0)$), which evolve freely under the evolution operator $U(\hat\epsilon_p,t) $, as defined in \eqref{eq:Ut}.
Still, expectation values are to be taken on a thermal state of the hamiltonian \eqref{Initial_diagonal}, which is diagonal in the operators $\eta_{i,p}$ (cfr. Eq.~\eqref{Initial_diagonal}).
Initial and final bosonic operators are related by the following sequence of transformations
\beq
\boldsymbol{b}_p (t)\quad
\stackrel{U(\hat\epsilon_p,t) }{\longrightarrow}\quad \boldsymbol{b}_p(0)\quad
\stackrel{B (\hat{\varphi}_p)}{\longrightarrow}\quad\boldsymbol{\eta}_{p}(0) \ ,
\eeq
equivalent to
\beq
\boldsymbol{b}_p(t)=U_{p}(t)B (\hat{\varphi}_p)\boldsymbol{\eta}_p(0) \ .
\eeq
These considerations allow us to write
\begin{multline} \label{thetaij}
 \langle\theta_{i}(x,t)\theta_{j}(y,t)\rangle=
\sum_{p,q\neq0}
e^{-i(px-qy)}
W^{\pm}_{\mu \nu }
\langle\boldsymbol{\eta}_{p,\mu}^{\dagger}(0)\boldsymbol{\eta}_{q,\nu}(0)\rangle ,
\end{multline}
where the sum over the dumb indices $\mu, \nu =\{ 1, 2, 3, 4\}$ is understood and the matrices $W^{\pm}$ have been defined in Eq.~\eqref{eq:Wpm}. Next, we observe that
\beq \label{etamunu}
 \langle\boldsymbol{\eta}_{p,\mu}^{\dagger}(0)\boldsymbol{\eta}_{q,\nu}(0)\rangle=
 \delta_{p,q} \left(F^{\beta}(p)_{\mu \nu} +(\delta_{\mu2}\delta_{\nu2}+\delta_{\mu4}\delta_{\nu4}) \right) \ ,
\eeq
where we further defined the matrix
\beq
F^{\beta}(p) =\text{diag}(f_{m}^{\beta},f_{m}^{\beta}, f_{0}^{\beta},f_{0}^{\beta})
\eeq
and $f_{k}^{\beta}\equiv f_{k}^{\beta}(p) = \frac{1}{e^{\beta \lambda_{k,p}}-1}$ ($k=m,0$) is the Bose function (and $\lambda_{k,p}$ in Eq.~\eqref{Eigenvalues}).
Finally, by using \eqref{etamunu} in Eq.~\eqref{thetaij}, and \eqref{thetapmxy} in Eq.~\eqref{eq:logCpm}, the exact expression of $C_{\pm}$ in Eq.~\eqref{General_corr} is easily obtained. The two matrix elements of  $W^{\pm}$ explicitly appearing in the final expression, can be evaluated directly from \eqref{eq:Wpm} and, in terms of the parameters $\hat{\varphi}_p$ defining the Bogoliubov transformation, they read
\begin{widetext}
\beq
\begin{array}{ll}
\displaystyle  \nonumber
W_{22}^\pm = &
\displaystyle
\Big\{
 \frac{\cos^2 \phi_p}{4 K_1}  (\cosh(2 \varphi_{1,p})-\cos(2 u_1 |p| t)\sinh(2\varphi_{1,p}))
 +  \frac{\sin^2 \phi_p}{4 K_2}  (\cosh(2(\Delta+\varphi_{1,p}))-\cos(2u_2 |p|t)\sinh(2(\Delta+\varphi_{1,p})))
 \\ \vspace{-0.2cm} \\
 & \displaystyle \pm \frac{\sin 2\phi_p}{4 \sqrt{ K_1 K_2}}  (- \cos((u_1- u_2)|p| t) \cosh(\Delta+2 \varphi_{1,p})+\cos((u_1+u_2)|p|t)\sinh(\Delta+2\varphi_{1,p}))
\Big\} \ ,
\end{array}
\eeq
\beq
\begin{array}{ll}  \label{W2244}
\displaystyle
W_{44}^\pm = &
\displaystyle
\Big\{
\frac{\cos^2 \phi_p }{4 K_2} (\cosh(2 \varphi_{2,p})-\cos(2 u_2|p| t)\sinh(2\varphi_{2,p}))
 +  \frac{\sin^2 \phi_p}{4 K_1}  (\cosh(2(\varphi_{2,p}-\Delta))-\cos(2u_2|p|t)\sinh(2(\varphi_{2,p}-\Delta)))
 \\ \vspace{-0.2cm} \\
 & \displaystyle \pm \frac{\sin 2\phi_p}{4 \sqrt{ K_1 K_2}}  ( \cos((u_1-u_2)|p|t) \cosh(2 \varphi_{2,p}-\Delta)-\cos((u_1+u_2)|p|t)\sinh(2\varphi_{2,p}-\Delta))
\Big\} \ .
\end{array}
\eeq
\end{widetext}

\section{Two-time correlations and FDT in the stationary state}\label{Appendix_FDT}

Here we study different Green's functions of system one and two after a thermal quench and we discuss their relation.
In particular the Keldysh, the retarded and the advanced Green's functions of system $i=1,2$ are defined, respectively, as follows
\beq
\begin{array}{l}
G_i^K(p,t_2,t_1) = \langle \{b_{i,p}(t_1), b_{i,p}^{\dag}(t_2) \} \rangle_{T_0} \ , 
\\ \vspace{-0.2cm} \\
G_i^{R}(p,t_2,t_1) = \theta(t_1-t_2) \langle [b_{i,p}(t_1), b_{i,p}^{\dag}(t_2) ] \rangle_{T_0} \ ,
\\ \vspace{-0.2cm} \\
G_i^{A}(p,t_2,t_1) = - \theta(t_2-t_1) \langle [b_{i,p}(t_1), b_{i,p}^{\dag}(t_2) ] \rangle_{T_0} \ ,
\end{array}
\eeq
where for completeness we consider the expectation value over a thermal state.
These functions turn out to be time translational invariant and depend only on the difference $t=t_1-t_2$,
immediately after the quench.
Moreover the response function (retarded correlator) does not depend on the initial
condition.
In particular, at the leading order in $p\to 0$ they read
\beq
\begin{array}{l}
\displaystyle
G_1^K(p,t) \simeq \frac{1}{2 u_1 |p|} e^{-i u_1 |p| t} \left[ m_1^{T_0} +  \left( \frac{u_2}{K_2} \frac{K}{u}  + \frac{K_1 u_1}{u K}\right) T_0 \right]
\\ \vspace{-0.2cm} \\
\displaystyle
G_2^K(p,t) \simeq \frac{1}{2 u_2 |p|} e^{-i u_2 |p| t} \left[ m_2^{T_0} +  \left( \frac{u_1}{K_1} \frac{K}{u}  + \frac{K_2 u_2}{u K}\right) T_0 \right]
\\ \vspace{-0.2cm} \\
\displaystyle
G_i^{R\backslash A}(p,t) =  \pm \theta(\pm t) \ e^{-i u_i |p| t} \quad \text{for $i=1,2$}
\end{array}
\eeq
with $m_i^{T_0}=m_i \ \text{cotanh}(m_0/2T_0)$.
Fourier transforming such functions in the frequency domain, one obtains
\beq\label{Eq_FDT}
G_i^K(p,\omega) = \frac{2 T^{T_0,\rm eff}_i}{\omega} \ ( G_i^R(p,\omega) - G_i^A(p,\omega)) \ ,
\eeq
with effective temperatures
\beq\label{T1_T2_T0}
\begin{array}{l}
\displaystyle
T_1^{T_0,\text{eff}}  =
%\frac{m_0}{4}  \cos\phi_0^2 =
\frac{m_1^{T_0}}{4} + \left(   \frac{u_2}{K_2} \frac{K}{u} + \frac{u_1 K_1}{u K} \right) \frac{T_0}{4} \simeq \langle \epsilon_{1,p} \rangle_{T_0}
\\ \vspace{-0.2cm} \\
\displaystyle
 T_2^{T_0,\text{eff}}  =
%\frac{m_0}{4} \sin\phi_0^2   =
 \frac{m_2^{T_0}}{4} +  \left(  \frac{u_1}{K_1} \frac{K}{u} +   \frac{u_2 K_2}{u K}  \right) \frac{T_0}{4} \simeq \langle \epsilon_{2,p} \rangle_{T_0} \ ,
\end{array}
\eeq
which are the generalization of (\ref{T1_T2}) to finite temperature quenches.
Eq. (\ref{Eq_FDT}) is the celebrated fluctuation-dissipation theorem (FDT) in the limit of
small frequencies (or classical limit) \cite{kamenev2011field}, which states a fundamental relation
between correlation and response functions in equilibrium systems.

\section{Leading analytic expressions of $C_{\pm}(x,t,T_0)$ after a thermal quench}\label{Appendix_thermal_quench}

In this section we provide a derivation of the equations that give the
leading order of the
correlation functions $C_{\pm}(x,t,T_0)$  and the
effective temperatures (\ref{Teffpm_T0}) and (\ref{Teff_T0}) after a thermal quench.

As we mentioned in the main text, Eq. (\ref{Eigenmodes_dynamics}) still holds, also at finite temperature.
The expectation values of the phase and density fluctuations at time $t=0$ however
are modified, in particular by the massless mode. These read
\beq\label{Eq_corr_theta_T0}
 \langle \theta_{i}(p,0) \theta_{j}(-p,0) \rangle
\simeq  %\frac{\pi u_1 u_2}{K_1 a^2 p^2} \sin\phi^2_0 T_0
\frac{\pi }{2 a^2 p^2} \frac{u_1 u_2}{K_1 K_2} \frac{K}{u} T_0
= \frac{1}{p^2}  \frac{\pi}{2 u K} T_0
\eeq
%\beq
% \langle \theta_{2}(p,0) \theta_{2}(-p,0) \rangle
%\simeq \frac{\pi u_2}{K_2 a^2 p^2} \cos\phi^2_0 T_0
%\eeq
%\beq
% \langle \theta_{1}(p,0) \theta_{2}(-p,0) \rangle
%\simeq \frac{\pi \sqrt{u_1 u_2}}{\sqrt{K_1 K_2} a^2 p^2}  \sin\phi_0 \cos\phi_0 T_0
%\eeq
%\beq
%\langle n_{1}(p,0) n_{1}(-p,0) \rangle
%\simeq  \frac{K_1}{2 \pi u_1} \left[  m_1\text{cotanh}\left(\frac{m_0}{2T_0}\right) + \frac{u_2 K}{K_2 u} T_0 \right]
%\eeq
%\beq
%\langle n_{2}(p,0) n_{2}(-p,0) \rangle
% \simeq  \frac{K_2}{2 \pi u_2} \left[  m_2\text{cotanh}\left(\frac{m_0}{2T_0}\right) + \frac{u_1 K}{K_1 u} T_0 \right]
%\eeq
\beq\label{Eq_corr_n_T0}
\begin{array}{c}
\displaystyle
\langle n_{i}(p,0) n_{j}(-p,0) \rangle
%\\ \vspace{-0.2cm} \\
\displaystyle
\simeq  \frac{1}{2\pi} \sqrt{\frac{K_i K_j}{u_i u_j}}  \left[  (-1)^{i+j} \sqrt{m_i^{T_0} m_j^{T_0}}  \right.
\\ \vspace{-0.2cm} \\
\displaystyle%\qquad\qquad\qquad\qquad
\left.
+ \sqrt{\frac{u_{k_1\neq i} u_{k_2\neq j}}{K_{k_1\neq i} K_{k_2 \neq j}}} \frac{K}{u} T_0 \right] \ ,
\end{array}
\eeq
with $m_i^{T_0} = m_i \text{cotanh}\left(\frac{m_0}{2T_0}\right)$ and $k_1,k_2=1,2$.
Therefore, in a thermal quench, both phase and density fluctuations contribute.
The building blocks (\ref{building_blocks}) become
\begin{widetext}
\beq
\begin{array}{ll}
\displaystyle
c_{ij} (x,t) \simeq
 \frac{1}{2} \int_0^{\infty} {\rm d} p \ e^{-\alpha^2 p^2} (1-\cos(p x)) \frac{1}{p^2}
 \displaystyle
  \Big\{  \Big[ (-1)^{i+j} \sqrt{\frac{m_i^{T_0} m_j^{T_0}}{K_i K_j u_i u_j}} +
  \frac{ T_0}{u K}\left(\frac{K^2}{K_1 K_2 } \sqrt{\frac{u_{k_1\neq i} u_{k_2\neq j}}{u_i u_j} } + 1\right) \Big]
   \cos((u_i-u_j)pt) -
   \\ \vspace{-0.2cm} \\
    \displaystyle
 \qquad\qquad\qquad
\Big[  (-1)^{i+j} \sqrt{\frac{m_i^{T_0} m_j^{T_0}}{K_i K_j u_i u_j}} +
\frac{ T_0}{u K}  \left( \frac{K^2}{K_1 K_2 } \sqrt{ \frac{u_{k_1\neq i} u_{k_2 \neq j}}{u_i u_j} } - 1 \right) \Big]
   \cos((u_i+u_j)pt) \Big\} \ .
 \end{array}
 \eeq
\end{widetext}
Note that this structure gives rise to the same light cones as for the quench from $T_0=0$.
From this we can read the final correlation length (in the case $u_1\neq u_2$)
\begin{widetext}
\beq
(\xi^{T_0}_Q)^{-1} =\frac{ \pi}{8} \Big[ \frac{m_0}{2} \frac{K}{u} \left( \frac{1}{K_1^2}+\frac{1}{K_2^2}  \right) \text{cotanh}\left( \frac{m_0}{2 T_0} \right) +
  \frac{ T_0}{u K}\left(\frac{K^2}{K_1 K_2 } \frac{u_1^2+u_2^2}{u_1 u_2}  + 2 \right) \Big] \ ,
  \eeq
\end{widetext}
which is compatible with the temperature (\ref{Teff_T0}).
Note that this expression has a simple interpretation in terms of
a two temperature system with temperatures given in (\ref{T1_T2_T0}), and generalizing Eqs. (\ref{T1_T2}) to a thermal quench.

In addition, the prethermal correlation length of the symmetric and the antisymmetric mode
(which can be deduced setting $u_1=u_2$ in the limit of large times) reads
\begin{widetext}
\beq
(\xi^{T_0}_{\pm})^{-1} = \frac{\pi}{8} \Big[ \frac{m_0}{2} \frac{K}{u} \left( \frac{1}{K_1^2}+\frac{1}{K_2^2}  \mp \frac{2}{K_1 K_2} \right) \text{cotanh}\left( \frac{m_0}{2 T_0} \right) +
  \frac{ T_0}{u K}\left(\frac{K^2}{K_1 K_2 } \frac{(u_1 \pm u_2)^2}{u_1 u_2}  + (2 \pm 2) \right) \Big] \ ,
  \eeq
which gives the effective temperatures (\ref{Teffpm_T0}).
\end{widetext}

\bibliographystyle{mioaps2}
\bibliography{dynamics_copuled_LL}

\end{document}